\documentclass[twocolumn]{aastex63}

\usepackage{amsmath}
\usepackage{multirow}

\newcommand{\masyr}{\,\mathrm{mas}\,\mathrm{yr}^{-1}}
\newcommand{\muw}{\mu_\mathrm{W}}
\newcommand{\mun}{\mu_\mathrm{N}}
\newcommand{\Msun}{\mathrm{M}_{\odot}}
\newcommand{\hst}{{\it HST}}
\newcommand{\gaia}{{\it Gaia}}
\newcommand{\vlos}{v_\mathrm{LOS}}
\newcommand{\vrad}{v_\mathrm{rad}}
\newcommand{\vtan}{v_\mathrm{tan}}
\newcommand{\vtot}{v_\mathrm{tot}}
\newcommand{\vx}{v_X}
\newcommand{\vy}{v_Y}
\newcommand{\vz}{v_Z}
\newcommand{\Mvir}{M_\mathrm{vir}}
\newcommand{\tperi}{t_\mathrm{peri}}
\newcommand{\tapo}{t_\mathrm{apo}}
\newcommand{\rperi}{r_\mathrm{peri}}
\newcommand{\rapo}{r_\mathrm{apo}}
\newcommand{\kpc}{\>{\rm kpc}}
\newcommand{\kms}{\>{\rm km}\,{\rm s}^{-1}}
\newcommand{\vesc}{v_\mathrm{esc}}
\renewcommand{\vec}[1]{\mathbf{#1}}

\received{May 26, 2020}
\revised{July 17, 2020}
\accepted{August 12, 2020}
\submitjournal{ApJ}

\shorttitle{\textit{HST} Proper Motions of NGC~147 and NGC~185}
\shortauthors{Sohn et al.}

\begin{document}

\title{\textbf{\Large \textit{HST} Proper Motions of NGC~147 and NGC~185: 
Orbital Histories and Tests of a Dynamically Coherent Andromeda Satellite Plane}}

\author[0000-0001-8368-0221]{Sangmo Tony Sohn} 
\affiliation{Space Telescope Science Institute, 
             3700 San Martin Drive, 
             Baltimore, MD 21218, USA}

\author[0000-0002-9820-1219]{Ekta Patel}
\affiliation{Department of Astronomy, University of California, Berkeley, 
             501 Campbell Hall, 
             Berkeley, CA, 94720, USA} 
\affiliation{Miller Institute for Basic Research in Science, 
             468 Donner Lab, 
             Berkeley, CA 94720, USA}

\author[0000-0003-4207-3788]{Mark A. Fardal}
\affiliation{Space Telescope Science Institute, 
             3700 San Martin Drive, 
             Baltimore, MD 21218, USA} 

\author[0000-0003-0715-2173]{Gurtina Besla}
\affiliation{Department of Astronomy, University of Arizona,
             933 North Cherry Avenue, 
             Tucson, AZ 85721, USA}
             
\author[0000-0001-7827-7825]{Roeland P. van der Marel}
\affiliation{Space Telescope Science Institute, 
             3700 San Martin Drive, 
             Baltimore, MD 21218, USA}
\affiliation{Center for Astrophysical Sciences, Department of Physics \& Astronomy, 
             Johns Hopkins University, 
             Baltimore, MD 21218, USA}

\author[0000-0002-7007-9725]{Marla Geha}
\affiliation{Department of Astronomy, Yale University,
             New Haven, CT 06520, USA}

\author[0000-0001-8867-4234]{Puragra Guhathakurta}
\affiliation{UCO/Lick Observatory, University of California Santa Cruz,
             1156 High Street, Santa Cruz, CA 95064, USA}

\begin{abstract}
We present the first proper motion (PM) measurements for the dwarf 
elliptical galaxies NGC\,147 and NGC\,185, two satellite galaxies of M31, 
using multi-epoch \hst\ imaging data with time baselines of $\sim 8$ years. 
For each galaxy, we take an error-weighted average of measurements from 
\hst\ ACS/WFC and WFC3/UVIS to determine the PMs. Our final results for 
the PMs are $(\muw, \mun)_\mathrm{N147} = (-0.0232, 0.0378) \pm (0.0143, 0.0146) 
\masyr$ for NGC~147, and $(\muw, \mun)_\mathrm{N185} = (-0.0242, 0.0058) \pm 
(0.0141, 0.0147) \masyr$ for NGC~185. The 2-dimensional direction of motion 
for NGC~147 about M31 is found to be aligned with its tidal tails. The 3-d 
positions and velocities of both galaxies are transformed into a common 
M31-centric coordinate system to study the detailed orbital histories of the 
combined M31+NGC~147+NGC~185 system via numerical orbit integration. We find that 
NGC~147 (NGC~185) had its closest passage to M31 0.3--0.5~Gyr ($\gtrsim 1.6$~Gyr) 
within the past 6~Gyr at distances of $\sim 70$~kpc (70--260~kpc). 
The pericentric times of NGC~147/NGC~185 correlate qualitatively well 
with the presence/absence of tidal tails seen around the galaxies. 
Our PMs show that the orbital poles of NGC~147, and also NGC~185 albeit to 
a lesser degree, agree within the uncertainties with the normal of the 
Great Plane of Andromeda (GPoA). These are the first measurements of the 
3-d angular momentum vector of any satellite identified as original GPoA members. 
Our results strengthen the hypothesis that the GPoA may be a dynamically 
coherent entity. We revisit previous claims that NGC~147 and NGC~185 are binary 
galaxies and conclude that it is very unlikely the two galaxies were ever 
gravitationally bound to each other. 
\end{abstract}


\section{Introduction}
\label{sec:intro}

Dynamical histories of satellite dwarf galaxies are essential for 
understanding their formation and evolution, as well as for exploring 
what roles they have played in the assembly of their hosts. Access to 
three-dimensional (3-d) space motions, which requires proper motion (PM) 
measurements combined with line-of-sight velocities ($\vlos$), allow 
reconstructing orbits of dwarf galaxies. In the past decade, PM results 
from both the \textit{Hubble Space Telescope} (\hst) and \gaia\ 
have tremendously advanced our understanding of the Milky Way (MW) halo. 
For example, PMs of satellite galaxies as a whole have allowed us to 
constrain the total mass of the MW by directly using 3-d motions without 
relying on model assumptions of the underlying velocity anisotropy 
distribution \citep{cal18,pat18,fri20}. 
Additionally, individual PMs of satellites have revealed interesting 
physics that has not been accessible through $\vlos$ alone.
Perhaps, the best examples are results showing that the Magellanic Clouds 
are on their first passage about the MW \citep{kal06,bes07,kal13},
and the first confirmation of a `satellites of satellites' hierarchy, 
i.e., clustering of low-mass dwarf satellites around the Clouds that 
entered the MW's halo as a group \citep{kal18,erk20,pat20}.
While the MW satellite system can be studied with this excellent 
level of detail, our general knowledge of satellite systems is limited 
primarily by the lack of equivalent 3-d motions for satellite populations 
around other MW-like galaxies. The M31 system provides the best opportunity 
for expanding our knowledge as well as for placing the MW system in a 
cosmological context since it is the only other MW-like system for which 
PM measurements are possible using current or planned space-based observatories. 

Compared to the MW, PM studies in the M31 system are in their very early 
stages. Among the M31 satellites, there are only two galaxies 
with PMs measured so far: M33 using VLBA and \gaia\ observations 
\citep{bru05,vdm19}; and IC~10 using VLBA observations \citep{bru07}.
While \gaia\ has proven invaluable for MW science, it has limited access 
to the M31 system since only the brightest stars in star-forming regions are 
detected at the distance of M31 \citep{vdm19}, whereas most M31 satellites 
are composed exclusively of old stars as revealed by their star formation 
histories \citep[SFHs;][]{wei14,ski17,wei19}.  

NGC~147 and NGC~185 are located in the outer halo of M31 and are the best 
observed of any dwarf elliptical (dE) galaxies, owing to their 
proximity to us at heliocentric distances of 724 and 636~kpc, 
respectively \citep{geh15}. Separated by only $\sim 1\degr$ on the sky, 
the two galaxies are fairly similar in mass, size, and stellar 
metallicity \citep{geh10}. Because it is considered an impressive 
coincidence to find two massive and similar satellites of M31 so 
close together on the sky with nearly identical line-of-sight 
velocities ($\vlos$), it has often been assumed NGC~147 and NGC~185 
form a dynamically bound pair \citep{vdB98}. 

However, the two galaxies show important differences. Based on the 
distances as measured by \citet[][which agrees very well with the 
measurements by \citealt{con12}]{geh15}, NGC~147 and NGC~185 are 
respectively located at distances of $724\pm27$ and $636\pm26$~kpc 
(M31-centric distances of $107$ and $160$~kpc; see Section~\ref{ss:3dposvel} 
for the details on the M31-centric coordinates), and the three-dimensional 
(3-d) separation between the two galaxies is $\sim 90$~kpc.
This indicates that NGC~147 and NGC~185 may be a chance alignment 
rather than a physical pair. If this physical separation is 
correct, their mutual gravity is unlikely to overcome the tidal force  
of M31 (i.e., their separation exceeds the Jacobi radius of the pair). 
Second, NGC~147 and NGC~185 exhibit stark differences in terms of their
stellar populations. By studying the color-magnitude diagram (CMD) 
down to the main sequence with \hst\ ACS/WFC, \citet{geh15} found that 
almost all of NGC~185's stars formed $>10$ Gyr ago, whereas NGC~147's stars 
formed on average much more recently, with half forming $\sim 6$ Gyr ago. 
Third, NGC~185 bears gas, while NGC~147 has a severe lack thereof 
\citep{you97,mar10}. Fourth, NGC~147 exhibits 50 kpc-long tidal tails 
pointing nearly radially toward and away from M31, while NGC~185 appears 
relatively undisturbed \citep{crn14}. Altogether, these properties suggest 
that the two galaxies may have had different orbital trajectories, although 
a scenario where they shared orbits until recently is not entirely ruled 
out \citep{ari16}. In this paper, we will reconstruct the mutual orbital 
histories of NGC~147 and NGC~185, providing the first assessment of 
these differences from a dynamical point of view.

NGC~147 and NGC~185 are important in the context of the formation of dEs. 
There are no morphological counterparts of these dE galaxies found in the 
MW halo, and so their formation mechanism may be related to e.g., 
environmental differences between the MW and M31. Interestingly, in 
hydrodynamic cosmological zoom-in simulations, \citet{gar19} successfully 
reproduced the majority of satellite population found in the Local Group, 
but they find that ``The simulations here do not produce any galaxies with 
densities as high as those of the baryon-dominated compact dEs around 
M31 \ldots'' Orbits of NGC~147 and NGC~185 may shed light on how such 
high-density dEs form and evolve in the halos of MW-like galaxies.

Lastly, \citet{iba13} found that more than half of M31's satellites 
rotate coherently within a narrow plane of satellites dubbed the `Great 
Plane of Andromeda' (GPoA). This structure is seen nearly edge-on from the 
MW allowing $\vlos$ to be used directly as a measure of whether the 
satellites exhibit coherent rotation. However, GPoA-like structures with 
subhalos exhibiting coherent $\vlos$ in cosmological simulations are rare 
\citep[][and references therein]{iba14,paw19} and only last for a 
short period of time, e.g., less than a typical satellite orbital timescale 
\citep{gil15,fer17,san20}. It is therefore crucial to understand whether 
the GPoA is a dynamically stable structure over time \citep{hod19}, 
or a transient structure. Both NGC~147 and NGC~185 are posited to be 
members of this plane, and thus provide the first opportunity to 
test this hypothesis.

In this paper, we present the first PM measurements for the M31 satellite 
dEs NGC~147 and NGC~185 using multi-epoch data obtained with \hst. 
This paper is organized as follows. In Section~\ref{s:pms}, we provide 
details on the data, the PM measurement process, and results of our 
measurements. In Section~\ref{s:spacemotions}, we derive the M31-centric 
space motions of NGC~147 and NGC~185, and use them to gain insights about 
M31's mass via escape velocities. In Section~\ref{s:orbhist}, we investigate 
the implications for the past orbits of NGC~147 and NGC~185 under various 
assumptions for the M31 mass and M31 tangential velocity ($\vtan$) zero 
point. In Section~\ref{s:gpoa}, we explore whether the 3-d motions of 
the satellites are aligned with the GPoA. Finally, in 
Section~\ref{s:conclusions}, we summarize the main results of our study.


\section{Proper Motions}
\label{s:pms}

\subsection{Data}
\label{ss:data}

%
\begin{deluxetable*}{lcccccccc}
\tablecaption{Observation summary of NGC~147 and NGC~185
              \label{t:obslog}}
\tablehead{
\colhead{}      &                   &                   & \multicolumn{3}{c}{{\bf Epoch~1}}                            & \colhead{} & \multicolumn{2}{c}{{\bf Epoch~2}}               \\
\cline{4-6}\cline{8-9}
\colhead{}      & \colhead{R.A. (J2000)}    & \colhead{Decl. (J2000)}   & \colhead{Date}    & \multicolumn{2}{c}{Exp. Time\tablenotemark{a}}      & \colhead{} & \colhead{Date}    & \colhead{Exp. Time\tablenotemark{a}}    \\
\colhead{Field} & \colhead{($\degr$)} & \colhead{($\degr$)} & \colhead{(Y-M-D)} & \multicolumn{2}{c}{(s$\times N$)}  & \colhead{} & \colhead{(Y-M-D)} & \colhead{(s$\times N$)} 
          }
\startdata
{\bf NGC~147}   &               &                 &            & F606W        & F814W      & &            & F606W     \\
ACS/WFC         & $8.207125$ & $+48.377528$ & 2009-11-21 & 1368s$\times$20   & 1373s$\times$32 & & 2017-11-18 & 1369s$\times$8 \\
WFC3/UVIS       & $8.295875$ & $+48.295556$ & 2009-11-21 & 1387s$\times$20   & 1392s$\times$32 & & 2017-11-18 & 1404s$\times$8 \\
\hline
{\bf NGC~185}   &               &                 &            & F606W        & F814W      & &            & F606W     \\
ACS/WFC         & $9.791042$ & $+48.441139$ & 2010-01-14 & 1358s$\times$20   & 1363s$\times$32 & & 2017-12-30 & 1369s$\times$8 \\
WFC3/UVIS       & $9.941250$ & $+48.424833$ & 2010-01-14 & 1375s$\times$20   & 1383s$\times$32 & & 2017-12-30 & 1404s$\times$8 \\
\enddata
\tablenotetext{a}{Exposure times in (seconds) $\times$ (number of exposures) format. 
                  Here we list the average of the entire exposures, 
                  but the actual individual exposure times vary by 
                  only a few percent in length.}
\end{deluxetable*}
%

%
\begin{figure*}
\gridline{ \fig{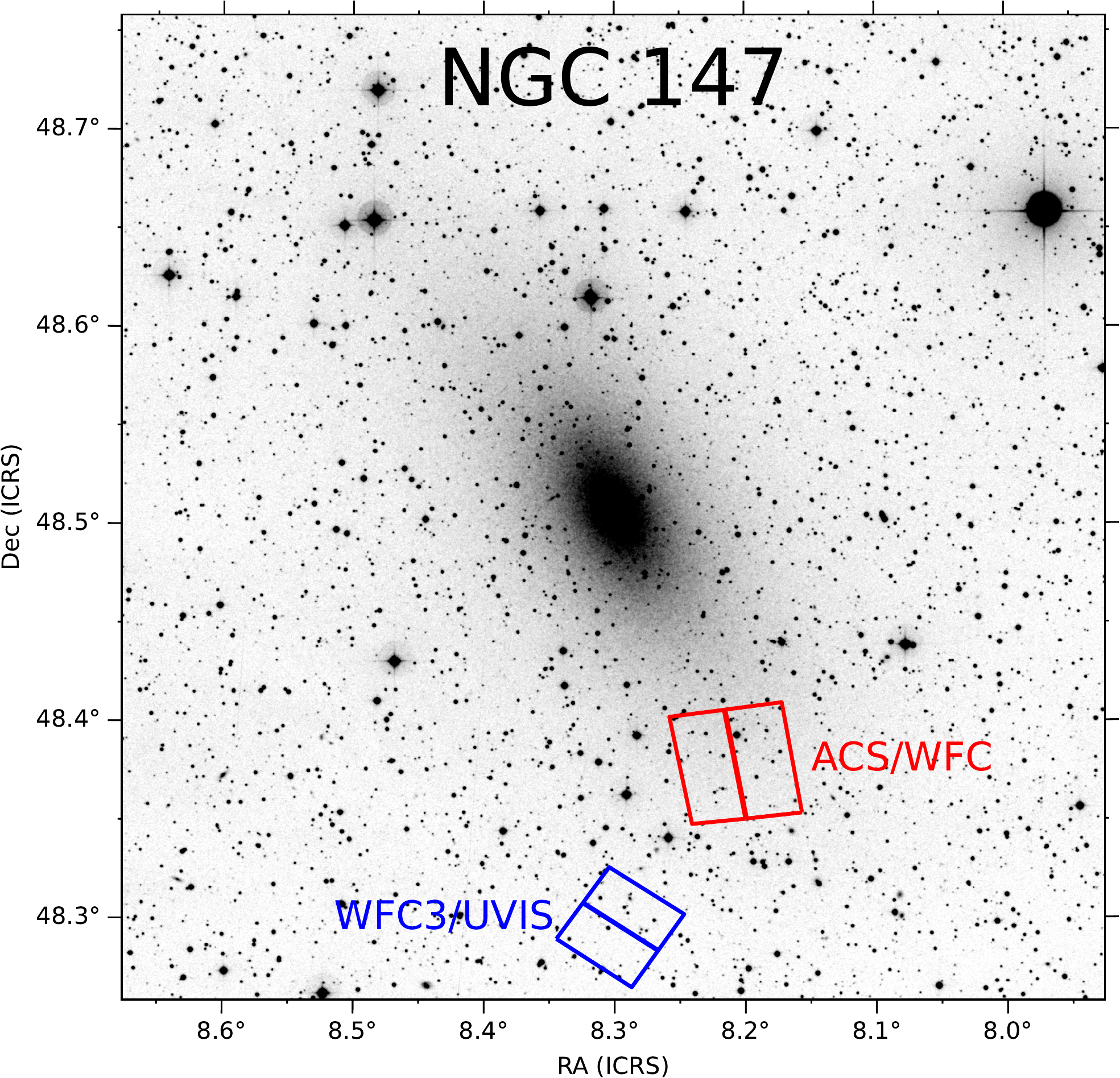}{0.5\textwidth}{}
           \fig{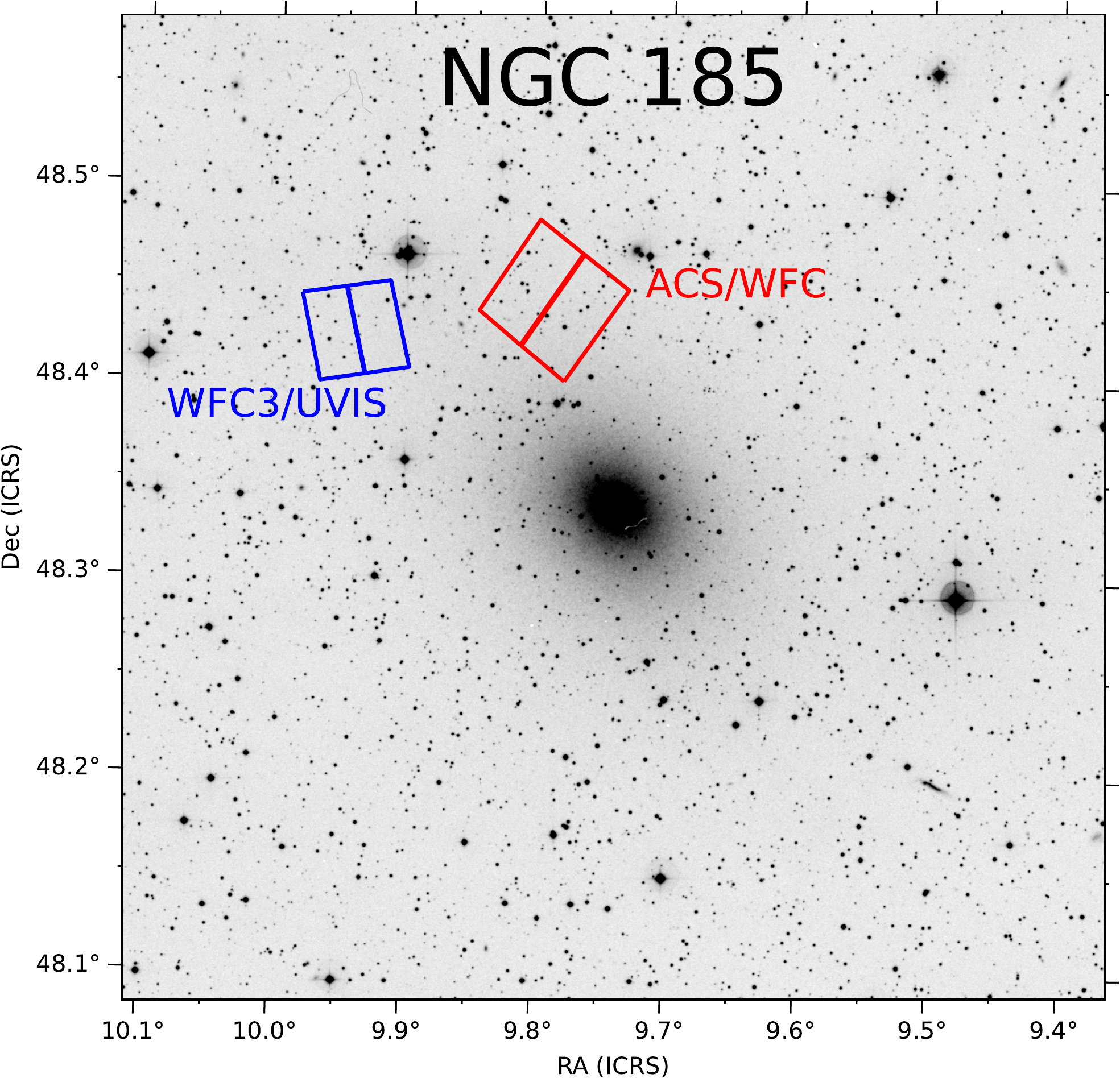}{0.5\textwidth}{}}
\caption{Field locations of our primary ACS/WFC (red) and parallel 
         WFC3/UVIS (blue) observations for NGC~147 (left panel) and 
         NGC~185 (right panel) plotted over a 30\arcmin$\times$30\arcmin\ 
         section of the sky centered on each galaxy from the STScI Digital 
         Sky Survey. The line that bisects each \hst\ field is a small 
         gap between the two CCDs; the CCD readout direction is roughly 
         perpendicular to this line. 
         \label{f:fields}}
\end{figure*}
%

We use \hst\ data obtained in two different epochs to measure the PMs 
of NGC~147 and NGC~185. For the first-epoch data of both galaxies, 
we use the images obtained through \hst\ program GO-11724 (PI: M. Geha). 
The ACS/WFC images for this program were used by \citet{geh15} to construct 
CMDs that reach several magnitudes below the main-sequence turnoff (MSTO) 
of the oldest populations, allowing them to study the SFHs of these galaxies 
in detail. The WFC3/UVIS images were obtained as parallel observations to 
the ACS/WFC images, and are located much farther from the center of 
each galaxy. Because of this, the stellar number densities in the WFC3/UVIS 
fields are far lower than those in the ACS/WFC fields. Both the ACS/WFC and 
WFC3/UVIS images were obtained using the broadband F606W and F814W filters. 
Figure~\ref{f:fields} shows the location of ACS/WFC and WFC3/UVIS fields 
on the sky with respect to NGC~147 and NGC~185.

All ACS/WFC and WFC3/UVIS fields were observed again $\sim 8$ years later
through \hst\ program GO-14769 (Co-PIs: S.~T. Sohn \& M.~A. Fardal) to 
measure the PMs presented in this analysis. We used the same telescope 
pointings and orientations as in the first-epoch observations to ensure 
that systematics related to the difference in positions on the detector 
(e.g., residual geometric distortions and charge transfer efficiency 
or CTE) are minimized. We obtained images only using the F606W filter 
in the second epoch since F606W gives the best astrometric handle 
on background galaxies based on our previous works \citep{soh12}.
All individual exposures of the first and second epochs were dithered 
using pre-defined patterns that maximize pixel-phase coverage.
All the data used in the study is available at 
\dataset[10.17909/t9-1pcz-se70]{https://doi.org/10.17909/t9-1pcz-se70}.

\subsection{Measurements}
\label{ss:measurements}

For the PM measurements of NGC~147 and NGC~185, we closely follow  
the established methods employed by our previous works on M31, 
Leo~I, and Draco/Sculptor dwarf spheroidal galaxies \citep{soh12,soh13,soh17}.  
Here we outline the general processes, but we refer readers interested in 
the details of the measurement process to the papers above.

All first- and second-epoch data are downloaded from the Mikulski 
Archive for Space Telescopes (MAST). We specifically download the 
{\tt \_flc.fits} images that have been processed for the imperfect 
charge transfer efficiency (CTE) using the correction algorithms of 
\citet{and10}. We determine a position and a flux for each star 
detected in the individual {\tt \_flc.fits} images using the FORTRAN 
code {\tt img2xym\_WFC.09x10} for ACS/WFC \citep{and06a} and an 
equivalent code for WFC3/UVIS. The measured positions are corrected 
for geometric distortion using the solutions provided by \citet{and06b} 
for ACS/WFC and by \citet{bel11} for WFC3/UVIS. For each galaxy, we then 
create high-resolution stacked images with pixel scales of 
$0.025$ and $0.020$ arcsec\,pix$^{-1}$ for ACS/WFC and WFC3/UVIS, 
respectively, using the first-epoch data.

Stars associated with NGC~147 and NGC~185 are identified via CMDs 
constructed from photometry of the first-epoch F606W and F814W images 
\citep{geh15}. Through this process, we select $> 30,000$ stars for 
each of NGC~147 and NGC~185 in their ACS/WFC fields. For the WFC3/UVIS 
fields, our final lists include $> 2,500$ stars for each of NGC~147 
and NGC~185. While there is about an order of magnitude difference 
compared to the numbers of stars in the ACS/WFC fields, the WFC3/UVIS 
samples are still large enough for measuring reliable PMs as will be 
shown in Section~\ref{ss:pmresults}.

Background galaxies are identified through a two-step process: first, 
an initial objective selection based on the quality-of-fit parameter 
output from the source detection code described above; second, 
visually inspecting each source identified in the initial stage, and 
adding additional extended sources that were missed by the source 
detection code. Consequently, we identify $\sim 170$ (NGC~147) and 
$\sim 220$ (NGC~185) useful background galaxies in the ACS/WFC fields, 
and $\sim 100$ (NGC~147) and $\sim 140$ (NGC~185) background galaxies 
in the WFC3/UVIS fields. The number ratios between background galaxies 
identified in the WFC3/UVIS and ACS/WFC fields correlate very well with 
the area ratio between the two detectors, $A_\mathrm{WFC3}/A_\mathrm{ACS} 
= 0.63$, indicating that by and large the number of background galaxies 
in these parts of the sky is isotropic.

For each star and background galaxy, we construct an empirical template 
by supersampling the scene extracted from the high-resolution stack. 
Each template constructed this way takes into account the PSF, the 
galaxy morphology (in case of the background galaxies), and the pixel 
binning. Templates are ``blotted'' back to each individual 
{\tt \_flc.fits} image to measure the position of each star 
or galaxy. For images of the first-epoch, templates are fit directly 
(since the templates were built {\it from} that epoch), while for the 
second-epoch, we include $7\times7$ pixel convolution kernels when 
fitting templates to allow for differences in PSF between the two epochs. 
These convolution kernels are derived by comparing PSFs of numerous 
stars associated with the target dwarf galaxies between the first- 
and second-epoch data. 

The reference frames are defined by averaging the template-based 
positions of stars from repeated exposures of the first epoch. 
We use the positions of stars in each second-epoch exposure to 
transform the template-measured positions of the galaxies into the 
reference frames. We then measure the difference between the first- 
and second-epoch position of each galaxy with respect to the stars 
associated with NGC~147 and NGC~185. For each galaxy, we calculate 
and apply a `local correction' using stars with similar brightness 
in the vicinity (typically $\pm 1$ magnitude, and within a 
200~pixel radius). This step ensures that any remaining systematics 
related to detector position or source brightness are corrected. 
For each individual second-epoch exposure, we take the 
error-weighted average over all displacements of background 
galaxies with respect to the NGC~147 and NGC~185 stars to obtain an 
independent PM estimate. The associated uncertainty is computed using 
the bootstrap method on the displacements of background galaxies. 
We note that the uncertainties in individual PM estimates are dominated 
by the background galaxy measurements, since there are far more stars 
in our fields than background galaxies, and since positions of stars are 
generally better determined than those of galaxies. The PM and associated 
error of each target galaxy are obtained by taking the error-weighted mean 
of the individual PM estimates. We then convert them to final PM 
results in units of $\masyr$ via multiplying by the pixel scale of our 
reference frames (0.050~arcsec\,pix$^{-1}$ for ACS/WFC and 
40~arcsec\,pix$^{-1}$ for WFC3/UVIS), and dividing by the time 
baseline of our observations. 

\subsection{Results}
\label{ss:pmresults}

%
\begin{deluxetable*}{lccccc}
\renewcommand{\arraystretch}{1.1}
\tablecaption{Proper Motion Results for NGC~147 and NGC~185.
              \label{t:pmresults}}
\tablehead{
\colhead{} & \multicolumn{2}{c}{\bf NGC~147}                                       & & \multicolumn{2}{c}{\bf NGC~185} \\
\cline{2-3} \cline{5-6}
\colhead{} & \colhead{$\muw$\tablenotemark{a}} & \colhead{$\mun$\tablenotemark{b}} & & \colhead{$\muw$\tablenotemark{a}} & \colhead{$\mun$\tablenotemark{b}} \\
\colhead{} & \colhead{($\masyr$)}              & \colhead{($\masyr$)}              & & \colhead{($\masyr$)}            & \colhead{($\masyr$)} 
}
\startdata
ACS/WFC          & $-0.0315\pm0.0172$ & \phs$0.0333\pm0.0177$ & & $-0.0272\pm0.0164$ &    $-0.0032\pm0.0168$ \\
WFC3/UVIS        & $-0.0047\pm0.0257$ & \phs$0.0473\pm0.0257$ & & $-0.0156\pm0.0276$ & \phs$0.0345\pm0.0301$ \\
\hline 
{\bf Weighted average} & $-0.0232\pm0.0143$ & \phs$0.0378\pm0.0146$ & & $-0.0242\pm0.0141$ & \phs$0.0058\pm0.0147$ \\
\enddata
\tablenotetext{a}{$\muw$ and $\mun$ are defined as the PMs in West 
                  ($\muw = - \mu_{\alpha}\cos\delta$) and North 
                  ($\mun = \mu_{\delta}$) directions, respectively.}
\end{deluxetable*}
%

%
%
\begin{figure*}[t]
\plottwo{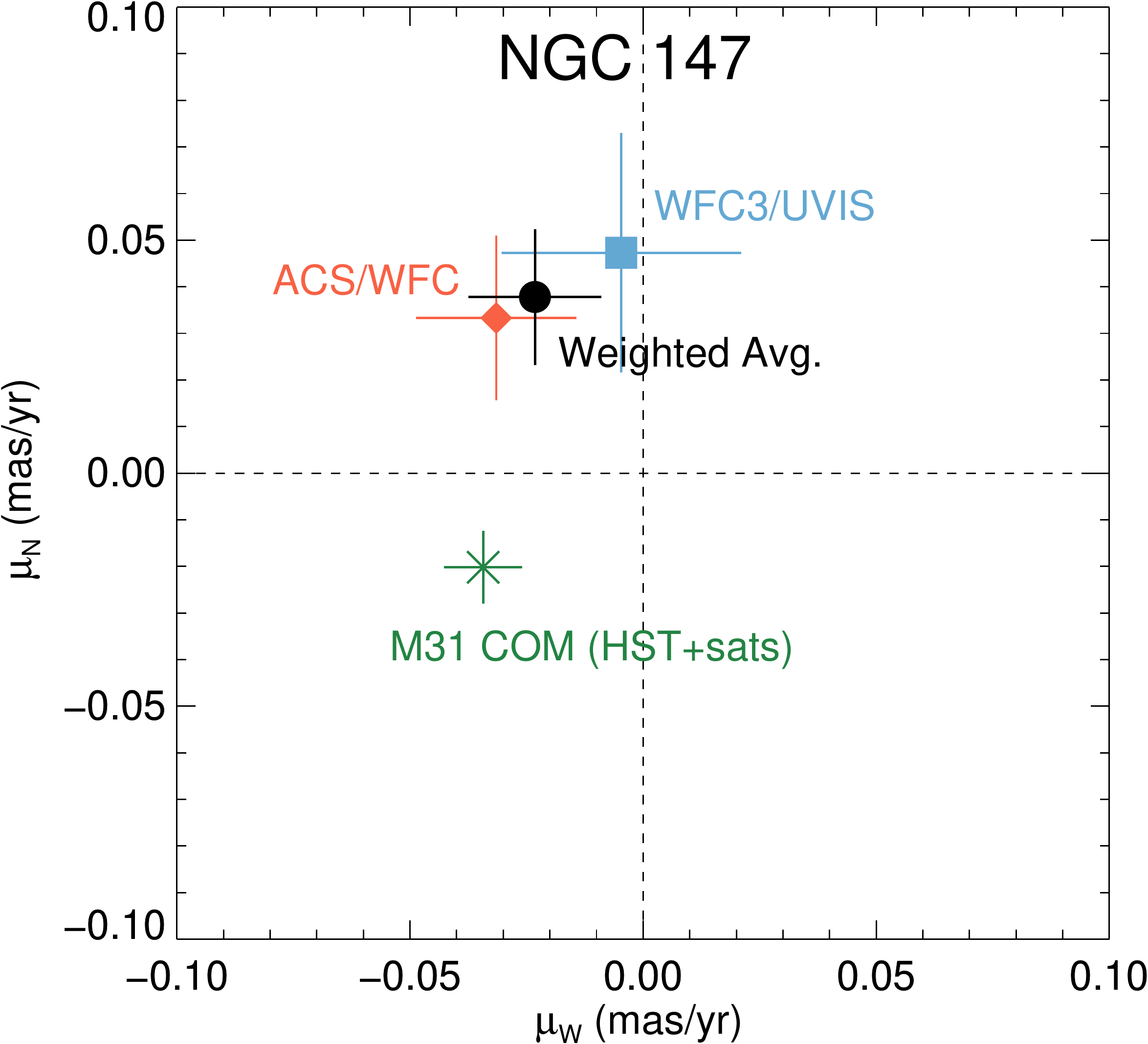}{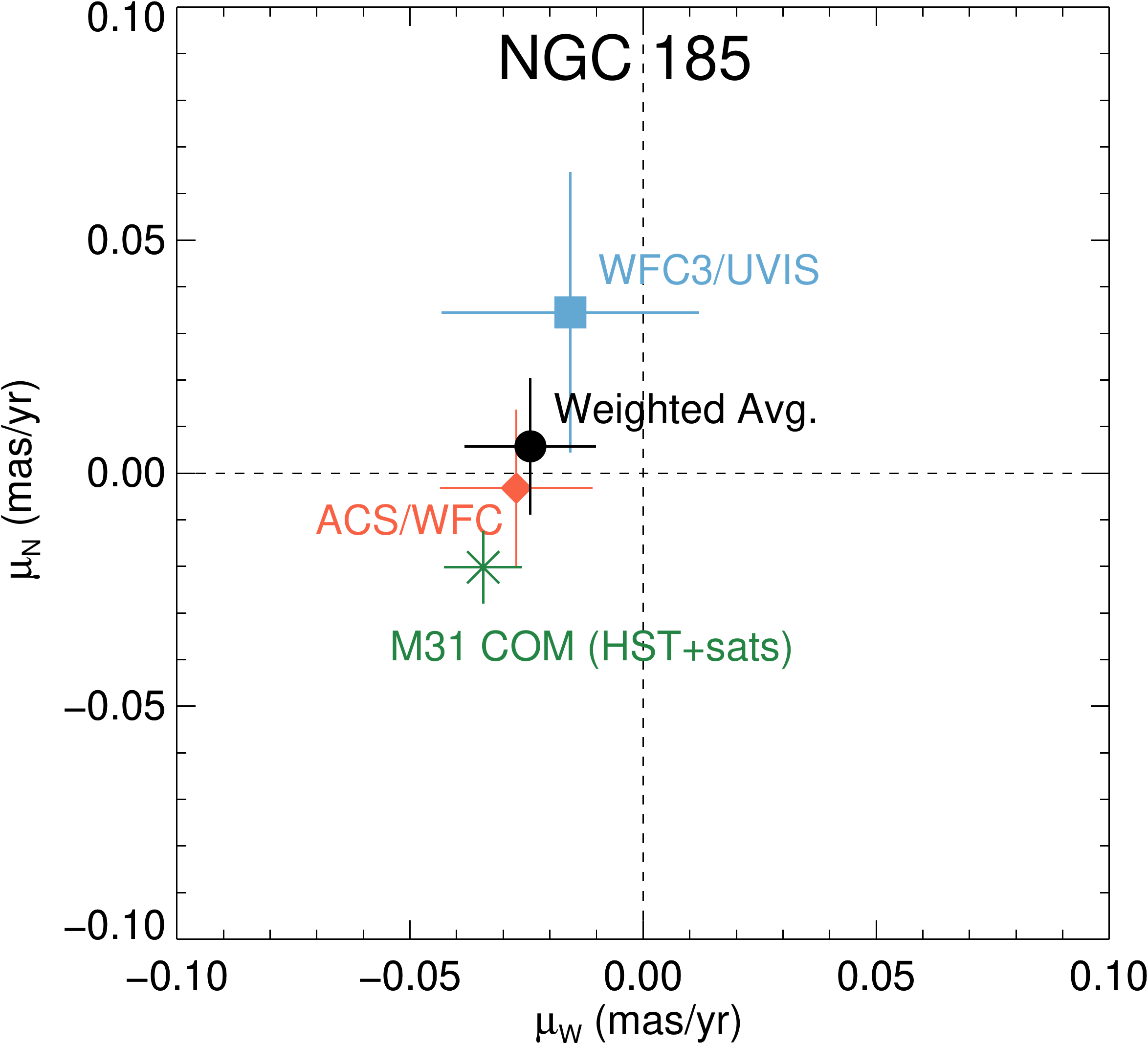}
\caption{Proper motion diagrams for NGC\,147 (left) and NGC\,185 (right).
         Light red and blue symbols are for the ACS/WFC and WFC3/UVIS 
         PM measurements, respectively. The weighted average of each 
         measurement is shown as the black dot. For comparison, we have 
         also plotted the M31 center of mass motion as determined in  
         \citet{vdm12} by taking the average of \hst\ measurements  
         \citep{soh12} and M31 motion estimate based on the kinematics 
         of its satellite galaxies. The dashed lines indicate 
         $\muw = \mun = 0$.
         \label{f:pmd}
         }
\end{figure*}

Table~\ref{t:pmresults} lists the PM results of the four target fields 
along with the error-weighted averages for each galaxy. The 
corresponding PM diagrams are shown in Figure~\ref{f:pmd}. 

The PM measurements for the ACS/WFC and WFC3/UVIS fields of both 
NGC~147 and NGC~185 appear to agree within their uncertainties in each 
coordinate. To test the statistical agreement among the measurements for 
different fields, we calculated the quantity 
\begin{equation}
\label{e:chisq}
  \chi^2 = \sum_{i} 
     \left[
     \left ( \frac{ \mu_{W,i} - {\overline \mu_{W}} }
               { \Delta \mu_{W,i} } \right )^2 
     +
     \left ( \frac{ \mu_{N,i} - {\overline \mu_{N}} }
               { \Delta \mu_{N,i} } \right )^2
     \right].
\end{equation}
This quantity is expected to follow a probability distribution 
with an expectation value of the number of degrees of freedom ($N_{\rm DF}$)
with a dispersion of $\sqrt{2N_{\rm DF}}$. Since we have two independent 
measurements each in two directions on the sky per galaxy, the $\chi^2$ is 
then expected to have a value of $2$ with an uncertainty of $2$. From 
Table~\ref{t:pmresults} and Equation~\ref{e:chisq}, we find $\chi^2 = 0.95$ 
and $1.33$ for NGC~147 and NGC~185, respectively, so we conclude that our 
results in Table~\ref{t:pmresults} are consistent within the quoted 
uncertainties. 

Because our target fields are offset from the centers of NGC~147 and 
NGC~185 (Figure~\ref{f:fields}), we consider the possibility that our 
measured PMs may not represent the center of mass (COM) motions of each 
galaxy. Any significant tangential (on-sky) motion of stars in our target 
fields relative to the center of each galaxy will affect the measured PMs. 
Here we consider the on-sky rotation (compared to the line-of-sight 
rotation) for each galaxy. 
While we are not able to directly measure the speed of on-sky rotation 
of either galaxy, the line-of-sight rotational motions provide hints 
on what to expect. \citet{geh10} measured the stellar $\vlos$ kinematics 
out to 7--9 times the effective radii ($r_\mathrm{eff}$) of NGC~147 and 
NGC~185. For both galaxies, significant lines-of-sight rotation signatures 
were detected at levels of 15--17~$\kms$. Our ACS/WFC target fields are in 
fact located at positions where stars with $\vlos$ measurements exist. 
From Tables~6 and 7 of \citet{geh10}, we selected stars that lie within 
the boundaries of our ACS/WFC fields and calculated the average heliocentric 
$\vlos$ with respect to the systemic velocities of each galaxy. Results are 
$-12.8~\kms$ and $+5.0~\kms$ for NGC~147 and 185, respectively. 
Given the apparent flattening of both galaxies on the sky, it is unlikely 
that their speed of on-sky rotation will be significantly higher than 
their line-of-sight rotation. We consider the extreme case of the on-sky 
rotation being equal to the line-of-sight rotations. At the distances 
of NGC~147 and 185, our final random PM uncertainties are 50 and 
43$\kms$, respectively. These are significantly larger than the rotation 
with respect to the COM motion even when the two components are combined. 
We therefore conclude that our measured PMs 
and associated uncertainties represent the COM motions of both NGC~147 
and NGC~185, and that no further corrections are required to the PMs as 
listed in the last rows of Table~\ref{t:pmresults}.

\section{Space Motions}
\label{s:spacemotions}

\subsection{Two-dimensional Motions of NGC~147 and NGC~185}
\label{ss:2dmotions}

%
\begin{figure}
\epsscale{1.15}
\plotone{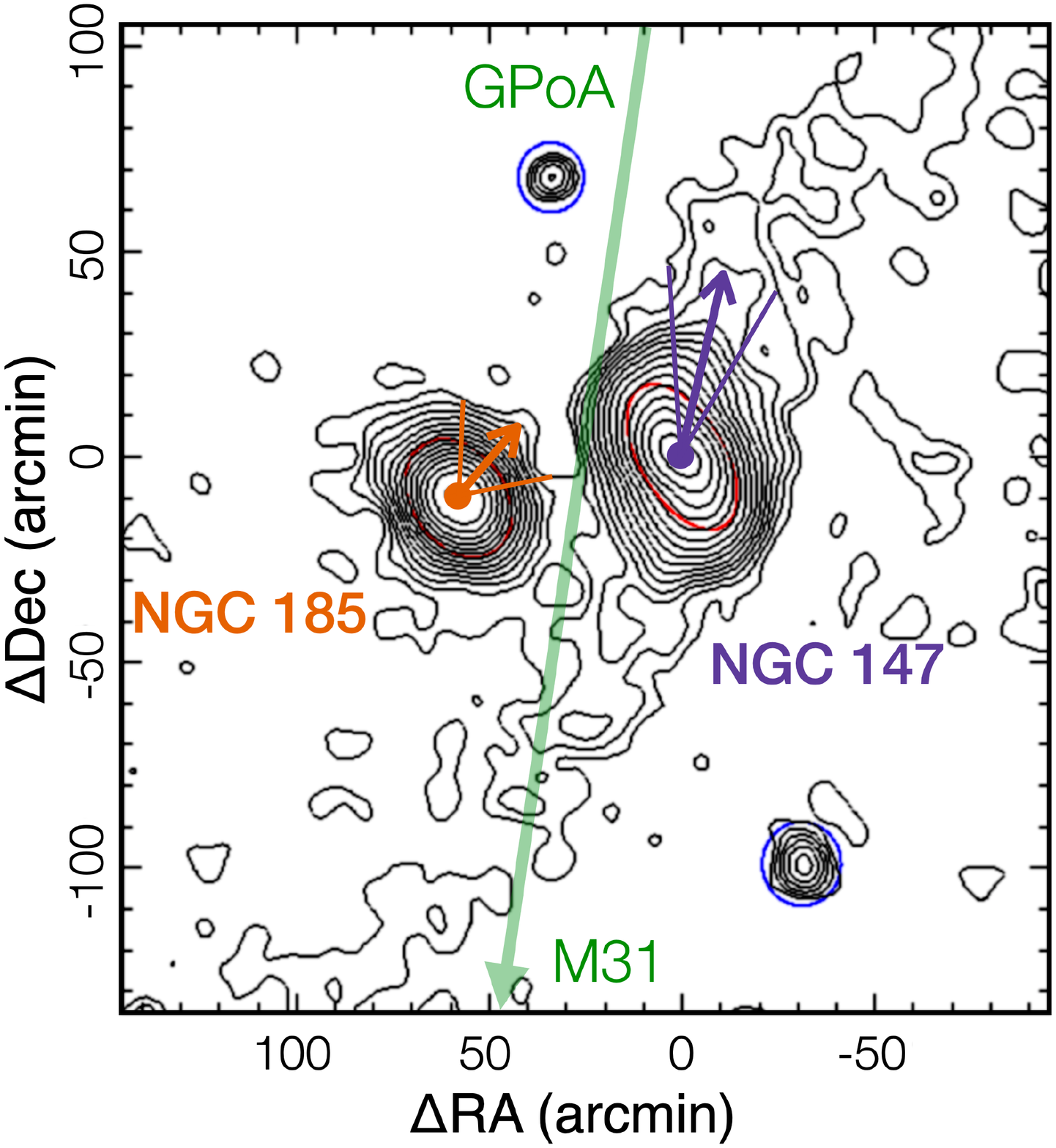}
\caption{Two-dimensional motions of NGC~147 (purple) and NGC~185 (orange) 
         with respect to M31 plotted over the surface density variation 
         of RGB stars by \citet{ari16}. The nominal tidal radii for 
         NGC~147 and NGC~185 are shown in red ellipses. The blue circles 
         on the top left and bottom right are Cas~II and And~XXV, 
         respectively. Details on the contours can be found in \citet{ari16}.
         M31’s PM contribution was subtracted from each dwarf galaxy's PM 
         result. The relative sizes of arrows were chosen to 
         represent the comparative 2-d motions of the two galaxies, 
         i.e., NGC~147 has a net 2-d motion about twice as fast as that 
         of NGC~185. The thin lines on either side of the arrows represent the 
         1$\sigma$ uncertainties of the 2-d motions based on the observed PM 
         uncertainties. The green tilted line that bisects the plot 
         represents the best fit to the GPoA discovered by \citet{iba13}. 
         M31 is located about $7\fdg4$ away from the origin of this plot 
         toward the direction indicated by the downward arrow of the green line. 
         \label{f:2dmotions}}
\end{figure}
%

With the measured PMs of NGC~147 and NGC~185, we first illustrate their
two-dimensional (2-d) motions with respect to M31 (i.e., projected 
on the plane of the sky). This is a useful exercise for visualizing how 
the two galaxies move relative to each other, and relative to other known 
structures such as the tidal tails around NGC~147. To do this, we subtract 
the PM contribution of M31 at each galaxy's distance and location on 
the sky to bring their 2-d motions into the M31 rest frame. We adopt the 
M31 PM estimate of \citet{vdm12} for this exercise.\footnote{Using the 
average of \hst\ and \gaia~DR2 measurements does not give much difference 
when visualizing, but we discuss the effect in Section~\ref{s:gpoa}.}
The resulting 2-d motions are shown in Figure~\ref{f:2dmotions}. 
The 2-d motion vector of NGC~147 is well aligned with its tidal tails 
surrounding the main body as expected for the case of an ongoing disruption, 
as well as with the direction of the GPoA. 
NGC~185's motion is not as well aligned, but is still marginally consistent 
with the direction of GPoA (angular offset between the green and orange 
lines is $\sim30$\degr). We will further discuss the 3-d motions of 
NGC~147 and NGC~185 relative to the GPoA in Section~\ref{s:gpoa}.

\subsection{Three-dimensional Positions and Velocities 
            in the M31-centric Rest Frame}
\label{ss:3dposvel}

To describe the space positions and velocities of NGC~147 and NGC~185, 
we follow a two-step process: first, we convert the observed parameters 
to Galactocentric positions and velocities; and second, we convert 
these to positions and velocities in the M31-centric rest frame.
For the first step, we calculate the current Galactocentric 
positions and velocities using measurements of observed parameters, 
including PM results from Section~\ref{ss:pmresults}, $\vlos$ from 
\citet{geh10}, and distances from \citet{geh15}. As with our previous studies 
\citep{soh12,soh13,soh17}, we adopt a Cartesian coordinate system 
$(X,\,Y,\,Z)_\mathrm{GC}$ with the origin at the Galactic center (thus 
denoted as the subscript GC), the $X$-axis pointing in the direction 
from the Sun to the Galactic center, the $Y$-axis pointing in the 
direction of the Sun's Galactic rotation, and the $Z$-axis pointing 
toward the Galactic north pole. The Galactocentric distance of the 
Sun and the circular velocity of the local standard of rest (LSR) are 
assumed to be $R_{0} = 8.29 \pm 0.16$~kpc, and $V_{0} = 239 \pm 5~\kms$, 
respectively following \citet{mcm11}. The solar peculiar velocity with 
respect to the LSR are adopted from the estimates of \citet{sch10}, i.e., 
$(U_\mathrm{pec},\,V_\mathrm{pec},\,W_\mathrm{pec}) = 
(11.10,\,12.24,\,7.25)~\kms$ with uncertainties of 
$(1.23,\,2.05,\,0.62)~\kms$. The positions and velocities of 
NGC~147 and NGC~185 in this frame are 
\begin{eqnarray}
  \label{e:r_147}
  {\vec r}_\mathrm{N147,GC} &=& (-357.2,\>608.8,\>-178.3) \kpc,\\
  \label{e:v_147}
  {\vec v}_\mathrm{N147,GC} &=& ( 11.9,\>71.1,\>174.6) \kms,
\end{eqnarray}
and 
\begin{eqnarray}
  \label{e:r_185}
  {\vec r}_\mathrm{N185,GC} &=& (-323.5,\>529.0,\>-158.1) \kpc,\\
  \label{e:v_185}
  {\vec v}_\mathrm{N185,GC} &=& ( 47.0,\>47.0,\>71.6) \kms,
\end{eqnarray}
respectively.

The conversion of these Galactocentric positions and velocities 
to the M31-centric ones involve subtracting M31's Galactocentric 
positions and velocities from Equations~\ref{e:r_147}--\ref{e:v_185} 
so that positions and velocities are as observed from the M31 center, 
and rotating the axes to make the $(X,\,Y,\,Z)_\mathrm{M31}$ 
specific to M31's orientation. We adopted M31's Galactocentric 
position from \citet{vdm12}, which assumes a heliocentric distance 
to the M31 of 770 kpc:
\begin{eqnarray}
  \label{e:r_m31}
  {\vec r}_\mathrm{M31,GC} &=& (-378.9,\>612.7,\>-283.1) \kpc.
\end{eqnarray}
Subtracting this from Equations~\ref{e:r_147} and \ref{e:r_185} 
gives the Cartesian coordinates shifted to the M31 center, but still 
oriented along the Galactocentric coordinate axes. To transform this 
into a coordinate system specific to M31's orientation, we adopted a 
system that has its origin at the center of M31, the $X$-$Y$ plane 
aligned with the disk of M31, $X$-axis pointing away from the (center 
of) MW, and $Z$-axis pointing normal to the $X$-$Y$ plane. This is 
the system adopted by \citet[][Appendix~B]{met07} with slight differences: 
we used a Galactocentric distance of the Sun of 8.29~kpc, heliocentric 
distance to M31 of 770~kpc, and inclination and position angle of the 
M31 disk of $i = 77\fdg5$ and $\theta = 39\fdg8$, respectively. The 
solution provided by \citet{met07}\footnote{We note that there is a 
crucial typo in one of the equations provided by \citet{met07}. The 
resulting matrix shown in Equation~B5 for $\mathbf{\mathrm{R}_{rpq}}$ 
should be transposed from its current form.} is for transforming the 
observed heliocentric coordinates directly to M31-centric ones. 
For our purpose, we independently derived a rotation matrix that 
transforms a coordinate system aligned with the Galactocentric 
$(X, Y, Z)_\mathrm{GC}$ but have its zero-point shifted to the center 
of M31 to a coordinate system specific to M31's orientation defined 
by \citet{met07}. The resulting matrix is
\begin{eqnarray}
\setlength\arraycolsep{3pt}
  \label{e:rotmatrix}
  {\vec R}_\mathrm{XYZ} = 
  \begin{pmatrix}
            -0.5859 & \phantom{-}0.6470 &           -0.4879 \\
            -0.7083 &           -0.1164 & \phantom{-}0.6962 \\
  \phantom{-}0.3937 & \phantom{-}0.7535 & \phantom{-}0.5265
  \end{pmatrix}
  .
\end{eqnarray}

The full transformation of a given satellite's Galactocentric 
position ${\vec r}_\mathrm{sat,GC}$ to the M31-centric position 
${\vec r}_\mathrm{sat,M31}$ is then given by
\begin{eqnarray}
  \label{e:rtransform}
  {\vec r}_\mathrm{sat,M31} = {\vec R_\mathrm{XYZ}}\,({\vec r}_\mathrm{sat,GC} - {\vec r}_\mathrm{M31,GC})^\mathrm{T}.
\end{eqnarray}
The Galactocentric velocity ${\vec v}_\mathrm{sat,GC}$ of a 
given satellite can be transformed to the M31-centric velocity 
using the same rotation matrix as follows:
\begin{eqnarray}
  \label{e:vtransform}
  {\vec v}_\mathrm{sat,M31} = {\vec R_\mathrm{XYZ}}\,({\vec v}_\mathrm{sat,GC} - {\vec v}_\mathrm{M31,GC})^\mathrm{T}.
\end{eqnarray}

The Galactocentric velocity of M31 (${\vec v}_\mathrm{M31,GC}$)  
in Equation~\ref{e:vtransform} is essential since it provides the 
zero point of the satellites' motions. 

The line-of-sight velocity ($\vlos$) of M31 has been known with great 
accuracy since the first measurement by \citet{sli13}, but {\em direct 
measurements} of tangential motion has only been available since 2012. 
Using multi-epoch \hst\ imaging data obtained using ACS/WFC and 
WFC3/UVIS, \citet{soh12} determined absolute PM of M31 stars in three 
fields. \citet{vdm12} corrected these measurements for internal 
kinematics, and combined them with the M31 motion estimate based 
on the kinematics of satellite galaxies of M31 and the Local Group. 
\footnote{\citet{sal16} also provided estimates of M31 $\vtan$ 
based on satellite kinematics. Their methodology and results are 
similar to those presented in \citet{vdm12}, so using the \citet{sal16}
results instead would not change Equation~\ref{e:vm31_hst} significantly.}
The resulting weighted-average estimate of the M31 COM tangential 
motion implies a Galactocentric velocity of
\begin{eqnarray}
  \label{e:vm31_hst}
  {\vec v}_\mathrm{M31, GC, \hst+sats} = (66.1, -76.3, 45.1) \kms.
\end{eqnarray}
A more recent estimate of M31's tangential motion was provided by 
\citet{vdm19}: their best estimate, after taking the average of the 
\hst\ and the \gaia~DR2 measurements implies
\begin{eqnarray}
  \label{e:vm31_hstdr2}
  {\vec v}_\mathrm{M31, GC, \hst+\gaia} = (35.0, -123.8, -17.0) \kms.
\end{eqnarray}
The difference in M31 3-d motions of these two measurements is $>80~\kms$ 
which is much larger than the uncertainties of NGC~147 and NGC~185's motions. 
Hence, the satellites' motions with respect to M31 will have a non-negligible 
systematic difference depending on which velocity zero point is adopted.
Due to its large distance, the M31 $\vtan$ is inherently difficult 
to determine, and systematic uncertainties for any single measurement or estimate 
is hard to quantify. We therefore believe the best approach is to average results 
from independent determinations instead of relying on one. 
Equations~\ref{e:vm31_hst} and \ref{e:vm31_hstdr2} were both determined 
by taking error-weighted averages of independent measurements or estimates, 
and thus considered the most reliable determinations of M31 $\vtan$ currently 
available. To explore the parameter space allowed by these two estimates of the 
M31 COM motion, and to examine the effects of using significantly different 
velocity zero points on the satellites' motions, we decide to consider 
both as velocity zero points for the M31 system and denote them 
as \hst+sats and \hst+\gaia\ (as in Equations~\ref{e:vm31_hst} and 
\ref{e:vm31_hstdr2}), respectively whenever necessary.

The positions and velocities of NGC~147 and NGC~185 in the M31-centric 
coordinate system are presented in Table~\ref{t:spacemotions}.
The uncertainties here and hereafter were obtained from a Monte Carlo 
scheme that propagates all observational uncertainties and their 
correlations, including uncertainties on the observed parameters 
adopted for the Sun. In the same table, we also list the M31-centric 
radial, tangential, and total velocities. 

%
\begin{deluxetable*}{lrrrcrrrrrr}
\tablecaption{Three-dimensional Positions and Velocities in the M31-centric Frame\tablenotemark{a}.\label{t:spacemotions}}
\tablehead{
\colhead{}       & \colhead{$X$}   & \colhead{$Y$}   & \colhead{$Z$}   & & \colhead{$\vx$}    & \colhead{$\vy$}    & \colhead{$\vz$}    & \colhead{$\vrad$}  & \colhead{$\vtan$}  & \colhead{$\vtot$}  \\
\colhead{Galaxy} & \colhead{(kpc)} & \colhead{(kpc)} & \colhead{(kpc)} & & \colhead{($\kms$)} & \colhead{($\kms$)} & \colhead{($\kms$)} & \colhead{($\kms$)} & \colhead{($\kms$)} & \colhead{($\kms$)}  
}
\startdata
\multicolumn{11}{c}{\hst+sats M31 $\vtan$} \\
\hline
\textbf{NGC~147} &  $-$66.4 & 58.0 & 60.8 & & 63.9$\pm$29.1  & 111.4$\pm$56.8 & 157.9$\pm$52.6  &   110.4$\pm$75.5 & 171.0$\pm$47.8 & 203.5$\pm$45.6 \\
\textbf{NGC~185} & $-$147.6 & 57.5 & 24.6 & & 78.0$\pm$27.8  &  17.6$\pm$51.2 &  99.3$\pm$47.6  & $-$50.3$\pm$44.8 & 117.2$\pm$40.8 & 127.5$\pm$31.8 \\ 
\hline
\multicolumn{11}{c}{\hst+\gaia\ M31 $\vtan$} \\
\hline
\textbf{NGC~147} &  $-$66.4 & 58.0 & 60.8 & & 46.2$\pm$36.4  & 127.1$\pm$62.0  & 238.6$\pm$56.7  &   175.8$\pm$61.5 & 210.6$\pm$39.0 & 274.3$\pm$54.1 \\ 
\textbf{NGC~185} & $-$147.6 & 57.5 & 24.6 & & 60.2$\pm$35.4  &  33.3$\pm$57.0  & 180.1$\pm$52.1  & $-$15.9$\pm$43.8 & 192.1$\pm$47.6 & 192.8$\pm$44.5
\enddata
\tablenotetext{a}{M31-centric frame is defined as having the origin 
at the center of M31, the $X$-$Y$ plane aligned with the disk, 
the $X$-axis pointing away from the MW, and the $Z$-axis pointing normal 
to the $X$-$Y$ plane.
}
\end{deluxetable*}
%

\subsection{Escape Velocity}
\label{ss:escapevel}

Since satellite galaxies act as tracers of their host galaxy's dark matter 
halo, measurements of their 3-d positions and velocities 
are powerful tools for constraining the total mass of their hosts. Under 
the general assumption that a satellite galaxy is {\em bound} to its 
host, the instantaneous total velocity $\vtot$ must be smaller than the 
escape velocity $\vesc = \sqrt{-2 \phi(r)}$, where $r$ is the distance from 
the center of host galaxy to the satellite, and $\phi(r)$ is the gravitational 
potential of the halo. Therefore, the escape velocity at the distance 
of a given satellite provides a lower limit for the total host galaxy mass 
required to keep the satellite bound. Here, we use the properties of 
NGC~147 and NGC~185, two satellites at relatively large 
galactic radii, to place constraints on the virial mass of M31.

\begin{figure}
\epsscale{1.2}
\plotone{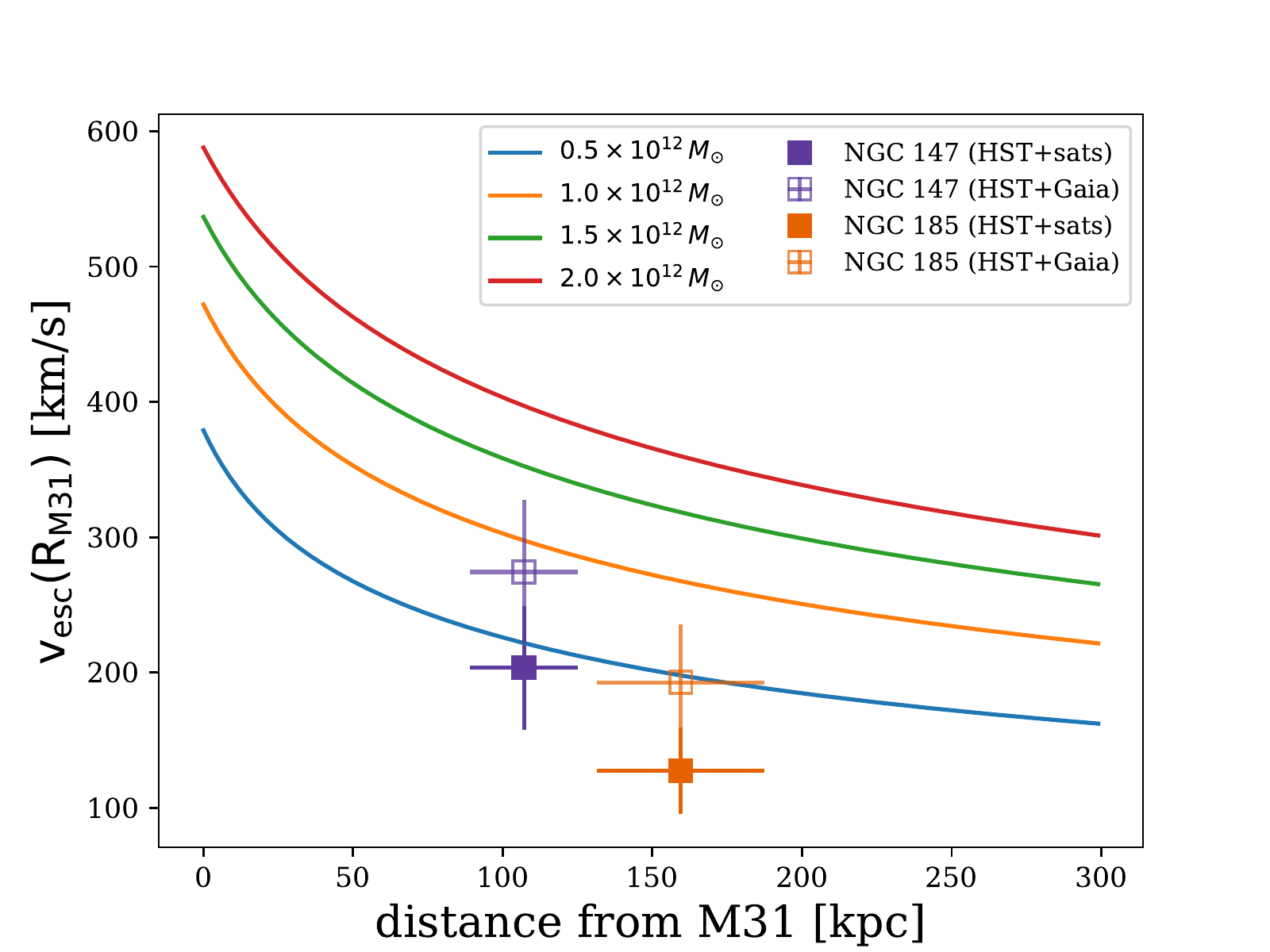}
\caption{The position and velocity of NGC~147 (purple) and NGC~185 
         (orange) using the \hst+sats M31 $\vtan$ (filled squares) 
         and the \hst+\gaia\ M31 $\vtan$ (open squares). M31 escape 
         velocity curves are also plotted as a function of radial 
         distance assuming an NFW profile. For both satellites to be 
         bound to M31, the minimal virial mass of M31 must be 
         $> 0.5\times10^{12}\,\Msun$ for the \hst+sats M31 $\vtan$ 
         and $> 10^{12}\,\Msun$ for the \hst+\gaia M31 $\vtan$.
         The green (low-mass) and red (high-mass) curves represent 
         the M31 masses adopted in our orbit calculations 
         (see Section \ref{s:orbhist}).
         \label{f:menc}}
\end{figure}

In Figure~\ref{f:menc}, we compare the observed total velocities of 
NGC~147 and NGC~185 taken from Table~\ref{t:spacemotions} to the 
escape velocities as a function of M31-centric distance for different 
cases of M31 virial masses ranging from $0.5$ to $2\times10^{12}~\Msun$. 
The escape velocity curves were calculated assuming an NFW profile for 
M31's halo and a concentration parameter calculated using Eq. 12 from 
\citet{pat17}. Square points indicate the present location of NGC~147 
(purple) and NGC ~185 (orange) and their $\vtot$ values computed using 
the \hst+sats M31 $\vtan$ (filled squares) and \hst+\gaia\ M31 $\vtan$ 
(open squares). Figure~\ref{f:menc} implies that both satellites are 
bound to M31 as they only require a minimal M31 virial mass of 
$\sim 0.5-1 \times \, 10^{12}~\Msun$ for $\vtot < \vesc(r)$. As more 
PMs become available for additional M31 satellites spanning the full 
virial extent of M31's halo, it will be possible to reconstruct M31's 
mass profile and derive more robust limits for M31's virial mass.

%
%

\section{The Orbital Histories of NGC~147 and NGC~185}
\label{s:orbhist}

\subsection{Orbit Integrations}
\label{ss:orbints}

To study the past orbital histories of NGC~147 and NGC~185, we 
numerically integrated their orbits backward in time using the current 
observed positions and velocities with respect to M31 as listed in 
Table~\ref{t:spacemotions}. The orbital integration scheme 
follows the methodology introduced in \citet{pat17}. 

In short, M31's potential is modeled as a static, axisymmetric, 
three-component model consisting of a stellar bulge, disk, and a 
dark matter halo. M31's position is not held fixed, thus it can move 
in response to the gravitational influence of the satellites. 
We consider two mass models for M31: a low mass M31 with a virial mass 
of $M_\mathrm{M31, vir} = 1.5\times10^{12}~\Msun$ and virial radius 
of 299 kpc, and a high mass M31 with $M_\mathrm{M31, vir} = 
2\times10^{12}~\Msun$ and a virial radius of 329 kpc. The bulge, disk, 
and halo parameters are adopted from Table~2 of \citet{pat17} for 
each mass model. The \citet{cha43} formula is adopted to account for 
the dynamical friction exerted on the satellites as they pass through 
the halo of M31.

Each satellite is modeled as a sphere following a Plummer density 
profile with a halo mass of $3\times 10^{10}~\Msun$. This halo mass 
is derived from the \citet{mos13} abundance matching relation using 
stellar masses adopted in \cite{gar19}. To determine the Plummer 
scale radius of each satellite, we adopt the measured, effective 
half-light radii and dynamical mass estimates reported in \citet{geh10}. 
We then find the scale radius where the Plummer mass enclosed at that 
distance matches the dynamical mass (i.e., the mass within the effective, 
half-light radius). This process yields Plummer scale radii of 1.5 kpc 
(NGC~147) and 0.9 kpc (NGC~185). Each satellite exerts forces on both 
M31 and the other satellite, thus these orbits consider the full 
3-body encounter between M31, NGC~147, and NGC~185.

Orbits were integrated backwards in time for a duration of 10~Gyr. 
Note that orbital results at times greater than 6 Gyr ago should be 
taken with caution since these models do not account for mass loss 
or growth \citep[see][for further justification]{pat17, pat20}. 
Furthermore, orbital uncertainties increase moving backwards in 
time (see Figure \ref{f:sep_vel}) and we have not accounted for any 
perturbations to the orbits of NGC~147 and NGC~185 owing to the 
gravitational influences of other massive M31 satellites, such as 
NGC~205, M32, or M33. To propagate the observed uncertainties, we 
randomly draw 2,000 samples of positions ($X, Y, Z$) and velocities 
($\vx, \vy, \vz$) to use as initial conditions for each satellite. 
In total, we integrate 8,000 total orbits to account for all four 
combinations of M31 virial mass and tangential velocity. These results are 
used to report the standard deviation of the distribution 
in each orbital parameter and are listed in Table \ref{t:orbparams}.

%
\begin{figure*}
\plotone{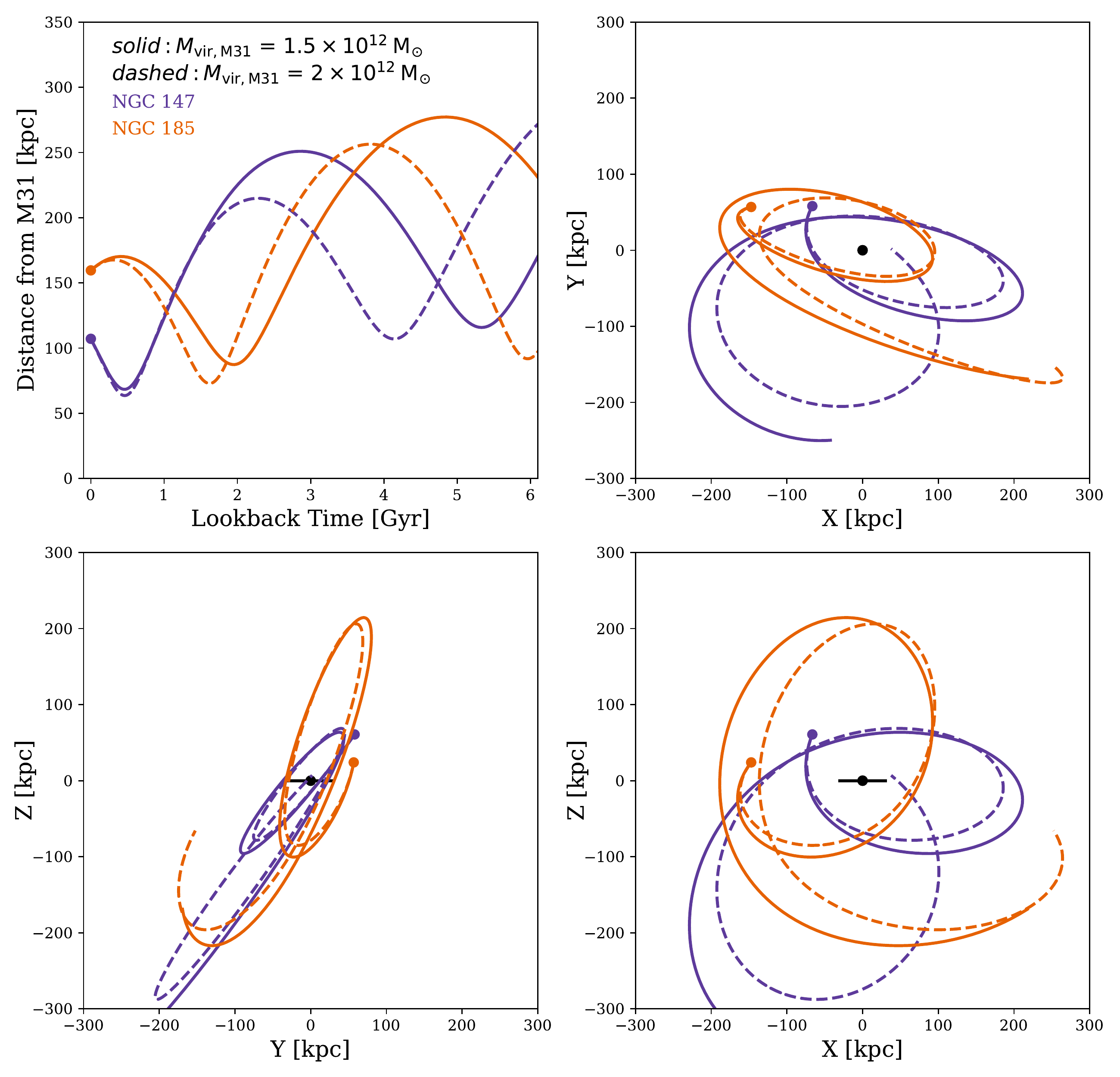}
\caption{Direct orbits of NGC~147 (purple) and NGC~185 (orange) in the 
         past 6~Gyr for a low-mass ($\Mvir = 1.5\times10^{12}\,\Msun$) 
         M31 model (solid lines) and a high-mass 
         ($\Mvir = 2\times10^{12}\,\Msun$) M31 model (dashed lines), 
         and using the \hst+sats M31 $\vtan$ as the velocity 
         zero point. The top left panel shows the separation between
         each galaxy and M31 as a function of time. In the top right, 
         bottom left, and bottom right panels, the orbital plane is 
         presented in the $X-Y$, $Y-Z$, and $X-Z$ planes of M31, 
         respectively. The current locations of NGC~147 and NGC~185 are 
         marked as orange and purple dots, respectively. The M31 center is 
         indicated as the black dot at $(X, Y, Z) = (0, 0, 0)$. 
         M31’s disk lies along the $Z$-plane in the M31-centric 
         frame as indicated by the black horizontal lines in the bottom 
         panels.
         \label{f:orbits1}}
\end{figure*}
%

%
\begin{figure*}
\plotone{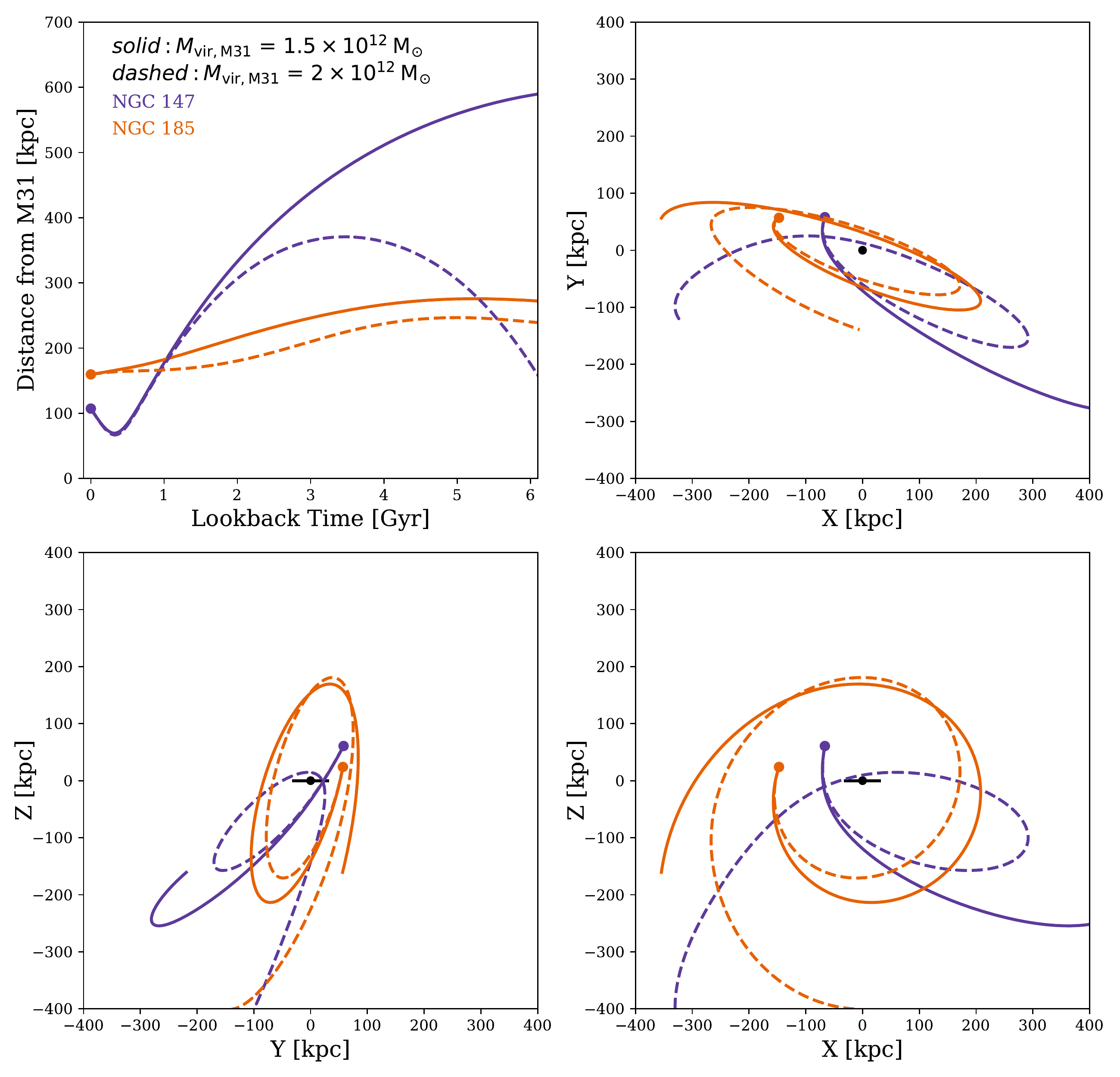}
\caption{Same as Figure~\ref{f:orbits1}, but using the 
         \hst+\gaia\ M31 $\vtan$ as the velocity zero point.
         Note that the axes limits are different from those in 
         Figure~\ref{f:orbits1}.
         \label{f:orbits2}}
\end{figure*}
%

\begin{deluxetable*}{cccccccc}
\setlength{\tabcolsep}{10pt}
\renewcommand{\arraystretch}{1.1}
\tablecaption{Orbital parameters.
              \label{t:orbparams}}
\tablehead{
\colhead{}       & \colhead{$M_\mathrm{vir,M31}$}   & \colhead{$f_\mathrm{peri}$\tablenotemark{a}}  & \colhead{$\tperi$\tablenotemark{b}} & \colhead{$\rperi$\tablenotemark{c}}  & \colhead{$f_\mathrm{apo}$\tablenotemark{a}} & \colhead{$\tapo$\tablenotemark{b}}  & \colhead{$\rapo$\tablenotemark{c}}   \\
\colhead{Galaxy} & \colhead{($\times 10^{12}~\Msun$)} & \colhead{(\%)}& \colhead{(Gyr)}             & \colhead{(kpc)}      &    \colhead{(\%)}    & \colhead{(Gyr)}            & \colhead{(kpc)}       }
\startdata
\multicolumn{8}{c}{\hst+sats M31 $\vtan$} \\
\hline
\multirow{2}{*}{\textbf{NGC~147}} & $1.5$ &  $99$ & $0.46\>[0.29, 0.64]$ & $68\>[42, 112]$ &  $89$ & $2.86\>[2.48, 4.99]$ & $251\>[193, 435]$ \\ 
                                  & $2.0$ & $100$ & $0.46\>[0.32, 0.68]$ & $64\>[39, 107]$ &  $97$ & $2.28\>[2.05, 3.47]$ & $215\>[177, 350]$ \\ 
\cline{2-8}
\multirow{2}{*}{\textbf{NGC~185}} & $1.5$ &  $96$ & $1.95\>[1.16, 3.42]$ & $87\>[47, 152]$ &  $98$ & $0.41\>[0.28, 3.74]$ & $170\>[152, 275]$ \\ 
                                  & $2.0$ &  $99$ & $1.62\>[1.09, 2.58]$ & $73\>[42, 135]$ & $100$ & $0.31\>[0.21, 2.62]$ & $168\>[150, 247]$ \\ 
\hline
\multicolumn{8}{c}{\hst+\gaia\ M31 $\vtan$} \\
\hline
\multirow{2}{*}{\textbf{NGC~147}} & $1.5$ &  $99$ & $0.32\>[0.20, 0.40]$ & $70\>[40, 110]$ & $44$ &               \nodata &          \nodata  \\ 
                                  & $2.0$ & $100$ & $0.33\>[0.22, 0.41]$ & $67\>[38, 109]$ & $71$ &  $3.47\>[2.47, 5.88]$ & $371\>[236, 570]$ \\
\cline{2-8}
\multirow{2}{*}{\textbf{NGC~185}} & $1.5$ &  $77$ &              \nodata &         \nodata & $73$ &  $5.23\>[0.75, 6.44]$ & $276\>[187, 433]$ \\ 
                                  & $2.0$ &  $90$ &              \nodata &         \nodata & $89$ &  $4.99\>[0.45, 5.22]$ & $247\>[176, 383]$ \\ 
\enddata
\tablecomments{The orbital parameters listed here are for the representative cases 
               shown in Figures~\ref{f:orbits1} and \ref{f:orbits2}. For uncertainties 
               of the orbital parameters, we quoted $[15.9, 84.1]$ percentiles 
               around the median of the distributions (see Appendix~\ref{s:app2} for details). 
               The parameters with missing data are for cases where the pericentric or 
               apocentric passage happened more than 6~Gyr ago, or will happen 
               in the future (negative Lookback time).}
\tablenotetext{a}{Fraction of orbits that made peri-/apo-centric passage in the past 10~Gyr.}
\tablenotetext{b}{Lookback time of the most recent peri-/apo-centric passage.}
\tablenotetext{c}{M31-centric distance at the most recent peri-/apo-centric passage.}
\end{deluxetable*}

\subsection{Orbital Analysis}
\label{ss:orbanalysis}
The direct orbits (orbits integrated using observed parameters listed 
in Table~\ref{t:spacemotions}) of the two satellites are shown in 
Figures~\ref{f:orbits1} and \ref{f:orbits2}. These orbits do not 
account for the measurement error on observed parameters, such as PM, 
$\vlos$, and distance. Table~\ref{t:orbparams} lists the peri- and 
apo-centric parameters resulting from the direct orbits.
We propagated the uncertainties in a Monte Carlo fashion as 
described in Section~\ref{ss:orbints}, and the correlation between these 
parameters and distributions of orbital parameters are presented in 
Appendices~\ref{s:app1} and \ref{s:app2}, respectively. 
In Table~\ref{t:orbparams}, we quoted the $[15.9, 84.1]$ percentiles 
around the median of the distributions (which corresponds to the 
$1\sigma$ or 68.2\%\ confidence intervals) as the uncertainties.

Overall, we find that the orbital parameters are strongly dependent 
on the velocity zero-point used for our orbital calculations. 
In particular, adopting the \hst+sats M31 $\vtan$ as the velocity 
zero point results in tighter orbits than adopting the \hst+\gaia\ M31 
$\vtan$ for both dwarf galaxies: i.e., the former leads to at least 
one full orbit around M31 (up to two full orbits for the higher-mass 
case) within 6 Gyr, while the latter barely results in one full orbit. 

There are, however, important similarities found among the four 
different cases shown in Figure~\ref{f:orbits1} and \ref{f:orbits2}.
First, as expected from the signs of radial velocities in 
Table~\ref{t:spacemotions}, we find for all cases that NGC~147 is 
currently moving away from M31, while NGC~185 is on an approaching orbit.
We also find that the pericentric distance and time of NGC~147 is very 
robust with small systematic differences regardless of which M31 mass 
is used, or which M31 $\vtan$ was adopted: i.e., NGC~147 had its 
pericentric approach to M31 0.3--0.5~Gyr ago at an M31-centric distance 
of 64--70~kpc on average. The pericentric time and distance for NGC~185, 
on the other hand, are not as well constrained, although we find that 
they are at significantly earlier times (1.5-2 Gyr ago for 
\hst+sats M31 $\vtan$ and $>6$ Gyr ago for the \hst+\gaia\ M31 
$\vtan$).  The apocenters, in contrast, are less well constrained for 
NGC~147 than for NGC~185, largely due to the different orbital phase 
at which we observe them. In the former the expected apocenter ranges 
from 215 to beyond 300~kpc depending on the choice of model, 
with the largest values coming for the reference frame and halo mass 
choice where the satellite is only weakly bound to M31. In the latter 
the apocenter ranges only from 168--276 kpc.

Interestingly, the pericentric times of NGC~147 and NGC~185 
correlate qualitatively very well with the presence and absence of 
tidal tails seen around the galaxies, respectively. In particular, 
the prominent tidal tails emanating from NGC~147 are likely still 
observable because of the recent (0.3--0.5~Gyr ago, but see below) 
pericentric approach to M31, whereas NGC~185 may have possessed tidal tails 
comparable to those of NGC~147 in the past (with the \hst+sats 
M31 $\vtan$ only), but have since dispersed and become significantly 
fainter in surface brightness. 
Future dynamical simulations of the two satellites using 
our PM results will help to refine these speculations and provide 
a more consistent physical picture of the tidal tails.

NGC~147 and NGC~185 are known to have contrasting differences regarding
their gas \citep{you97,mar10} and stellar contents \citep{geh15}. 
NGC~147 is devoid of gas, and in addition to the ancient stars 
($> 12$~Gyr old) that make up $\sim 40$\% of its stellar population, 
more than half of its stars were formed 5--7~Gyr ago. NGC~185 on the 
other hand, contains some gas and dust with the bulk of the stars being 
formed 8--12.5~Gyr ago. While the characteristics of our orbits are too 
sensitive to observational and modeling uncertainties to draw firm 
conclusions about their relationship to the stellar population (cf. 
Figures~\ref{f:orbits1} and \ref{f:orbits2}), NGC~147's orbit is 
likely more radial and higher-energy than that of NGC~185, with both 
a larger apocenter and a smaller pericenter. This suggests that 
NGC~185 may have experienced a longer, more gentle period of 
evolution within M31's halo, with its stellar population quenched 
more by slow starvation than ram pressure stripping. By contrast, 
the large population of intermediate-age stars in NGC~147 may have 
been induced by tidal forces owing to M31 at an earlier pericentric 
passage when it still retained a large amount of gas. Viewed more 
broadly, our results show that the two galaxies are likely to be 
orbiting independently, and therefore it is not a surprise that their 
gas properties and SFHs are significantly different.

\subsection{Are NGC~147 and NGC~185 a Bound Pair?}
\label{ss:bound}

Our orbital results also provide insights into the possibility of past 
interactions between the two dwarfs. This is important in the context 
of previous claims that NGC~147 and NGC~185 may be `binary galaxies', 
i.e., born as a group, and bound to each other while orbiting around M31 
\citep{vdB98,geh10}. \citet{geh10} calculated the energy criterion 
$b\equiv (2GM_{sys})/(r\Delta v^2)$, which when $>1$ implies that the 
potential energy between the two satellites is greater than their 
kinetic energy. In other words, when the satellites have $b>1$, they 
are strongly interacting with one another. In \citet{geh10}, a value 
of $b=1.6\pm1.1$ was calculated using the line-of-sight distance between 
the two satellites, the difference in their $\vlos$, and a system mass 
comprised of the sum of their dynamical masses. Here, we use the 
difference in their 3-d position vectors ($\sim$89 kpc), their total 
relative velocities ($\sim112 \kms$), and the sum of their halo masses 
($6 \times 10^{10}\,M_{\odot})$ to find that $b=0.47^{+0.51}_{-0.43}$, 
where these values represent the median and the extents of the 25th 
and 75th quartiles of the resulting $b$ values using all 2,000 MC 
drawings, accounting for measurement errors in all relevant 
observational parameters. We conclude that NGC~147 and NGC~185 are not 
strongly interacting with one another at present. A combined mass of 
at least $\sim1.3\times 10^{11}\, M_{\odot}$, about twice the combined 
mass considered here, is required to reach $b=1$ using the current 
separation and relative velocity for NGC~147 and NGC~185.
 
\begin{figure}
\epsscale{1.2}
\plotone{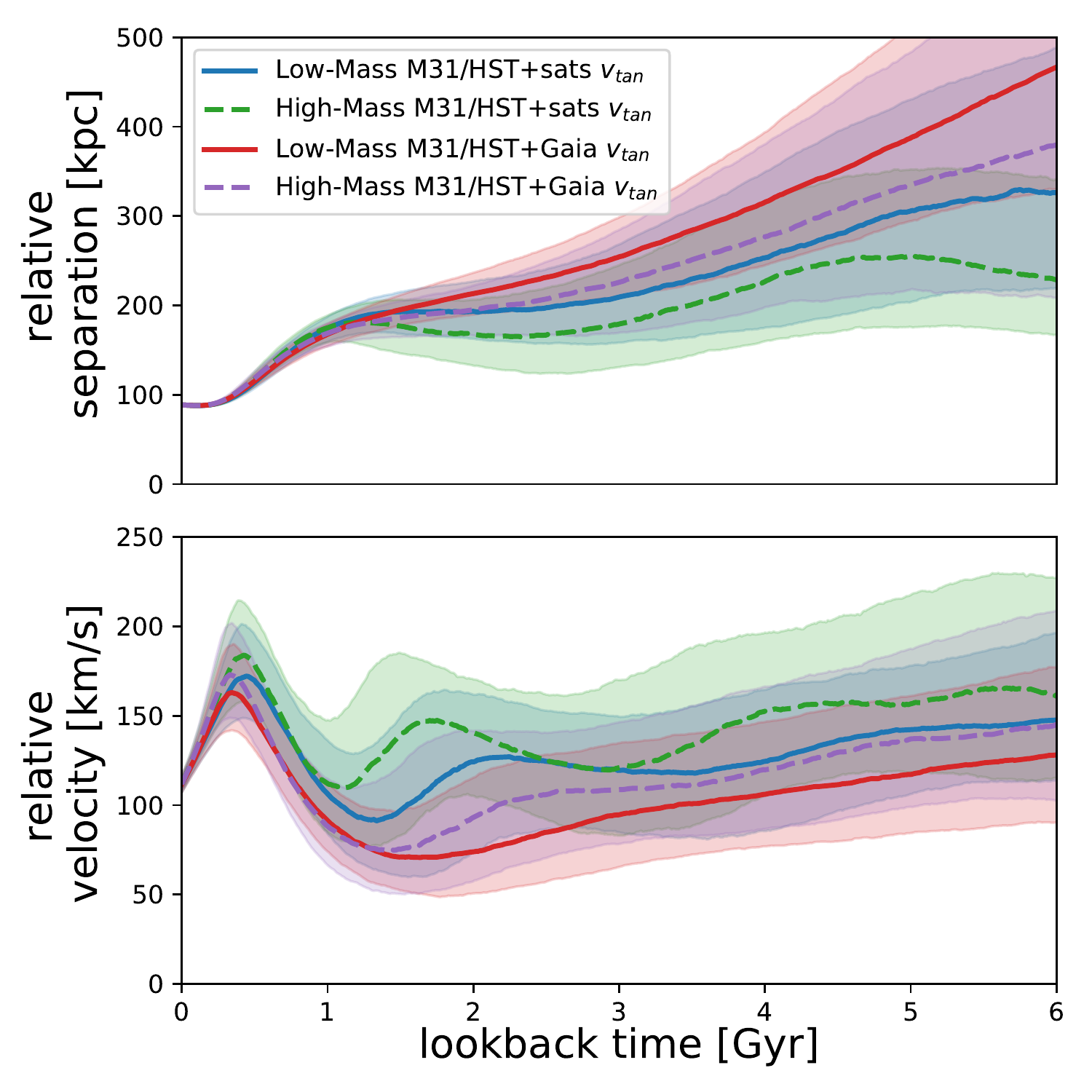}
\caption{The top panel shows the relative separation of NGC~147 and 
         NGC~185 over the last 6 Gyr. Results are shown for each 
         combination of low (solid lines) and high (dashed lines) M31 
         mass and tangential velocity. For each combination, lines 
         indicate the median separation between the two satellites and 
         shaded regions correspond to the 25th and 75th quartiles of 
         the distribution in separation at each time across all 2,000 
         orbit calculations. The bottom panel shows the corresponding 
         total relative velocity between NGC~147 and NGC~185 computed 
         in a similar fashion. The minimum separation 
         between the two galaxies is $\sim$100 kpc and occurs at 
         relatively recent times ($<$ 0.5 Gyr ago) for all results. 
         \label{f:sep_vel}}
\end{figure}

With measured PMs, we can also evaluate whether the satellites were 
a bound pair in the past. Figure~\ref{f:sep_vel} shows the median 
distance and total relative velocity of NGC~147 and NGC~185 over the 
last 6~Gyr. These quantities were calculated by taking each set of 
2,000 orbits per M31 mass and tangential velocity value, computing 
the relative orbits of NGC~147 and NGC~185, and then finding the 
median and quartiles (shaded regions) for the separation and velocity. 
As their minimum separation of $\sim100$ kpc is only achieved in the 
last 0.5~Gyr, we conclude that the satellites were not a bound pair 
over the last 6~Gyr and therefore the tidal tails around NGC~147 are 
likely a result of NGC~147's recent pericentric passage around M31 at 
0.3--0.5~Gyr ago. Furthermore, the satellites likely entered the halo 
of M31 individually rather than during a group infall, in contrast to 
the Magellanic Clouds and their associated satellites 
\citep[e.g.][]{kal18, erk20, pat20}.

Our analysis above apply to the typical behavior of our orbit models. 
A small fraction of our orbits, however, do allow approaches that are 
much closer than the present-day distances of the two galaxies. For 
instance, over our combinations of mass and velocity zero point, roughly 
2\%\ of the orbits allow approaches closer than $\sim 30 \kpc$, a 
distance at which one could contemplate the two galaxies being bound. 
In most cases these close approaches occur on previous encounters, 
$> 1.5 Gyr$ into the past. Thus we do not entirely rule out that the 
two galaxies were previously bound, though for the combinations of 
observational uncertainties, M31 mass, and M31 $\vtan$ adopted in this 
analysis we find it highly unlikely.

%
%

\section{The Great Plane of Andromeda}
\label{s:gpoa}

Both NGC~147 and NGC~185 are considered to be members of the GPoA 
\citep{iba13,con13}. In Section~\ref{ss:2dmotions}, we have already 
shown that the 2-d motion of NGC~147 is consistent with evidencing 
rotation within the GPoA, while NGC~185's direction of motion is 
somewhat tilted about it. Here, we use the 3-d motions of NGC~147 
and NGC~185 based on our PM measurements to provide clues to whether 
the satellites are members of the GPoA both today and as a function 
of time. We also provide useful arguments on whether the GPoA is a 
long-lived dynamical structure or not using our PM measurements.

\subsection{Orbital Poles of M31 Satellites Compared to the GPoA Normal}
\label{ss:orbpole}

%
\begin{figure*}
\epsscale{1.0}
\plotone{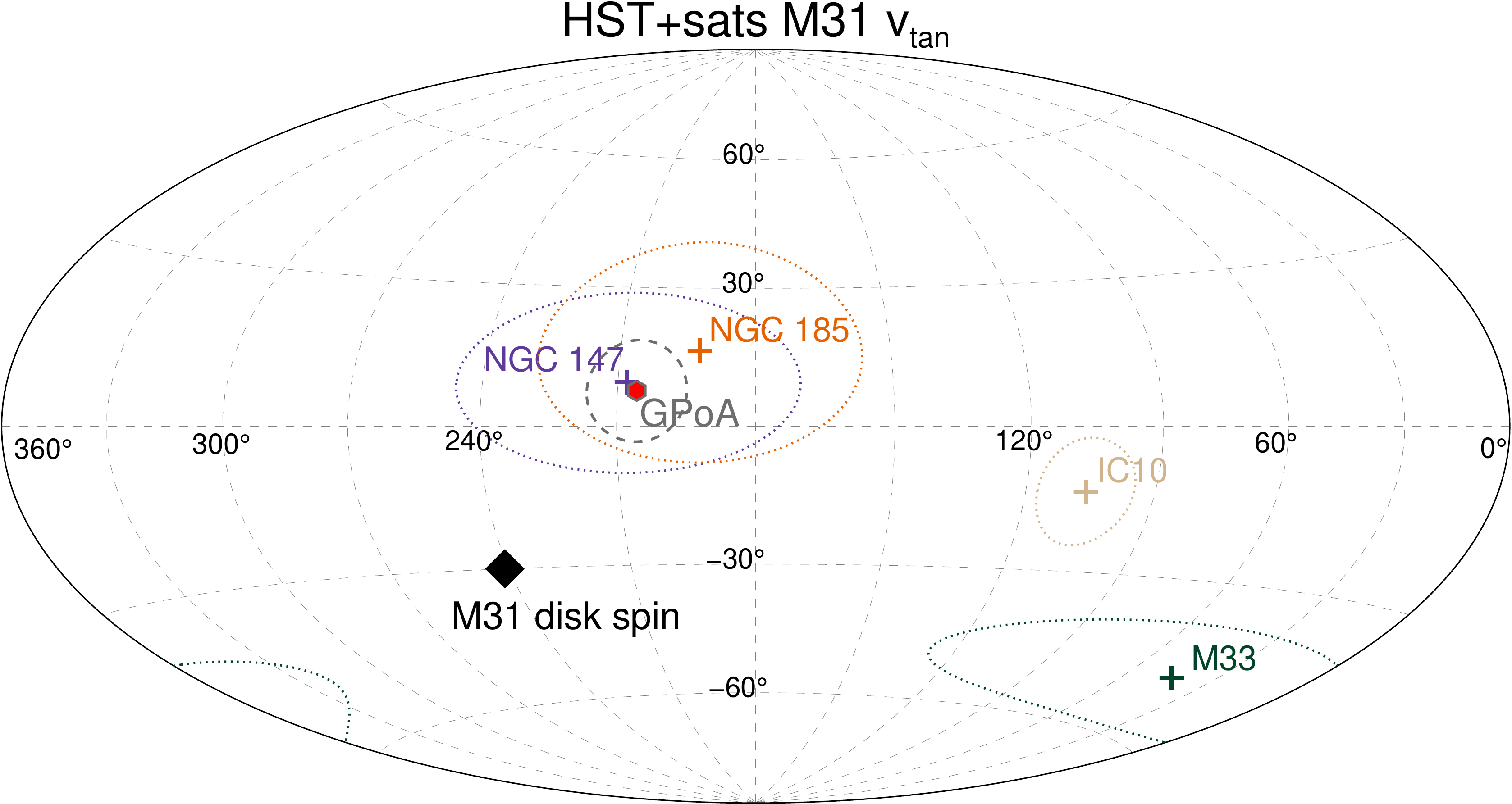}
\caption{Orbital poles of M31 satellites with existing proper motion 
measurements in Galactic coordinate compared to the normal direction 
of the Great Plane of Andromeda (GPoA) shown as the red hexagon at 
$(l, b) = (205\fdg8, 7\fdg6)$ from \citet{paw13}. The circle with 
gray dash line centered on the red hexagon represents the rms width of 
the GPoA after fitting the plane to the 19 satellites as defined in 
\citet{paw13}. For each satellite, we plot the orbital pole as plus signs 
along with the $1\sigma$ ($\simeq 68$\%) confidence region as an ellipse 
with dotted line. For M33, distance and line-of-sight velocity were 
adopted from \citet{mcc12}, and proper motion from \citet{bru05}. 
For IC~10, distance was adopted from \citet{mcq17}, line-of-sight 
velocity from \citet{huc99}, and proper motion from \citet{bru07}.
This plot assumes using the \hst+sats M31 $\vtan$ as the velocity 
zero point. The corresponding plot using \hst+\gaia\ M31 $\vtan$ 
is essentially identical to this one considering uncertainties, 
so we chose not to include it here.
\label{f:orbpoles}}
\end{figure*}
%

In Figure~\ref{f:orbpoles}, we plot the present day orbital poles of all 
M31 satellites with measured PMs, i.e., M33, IC~10, NGC~147, and NGC~185.
As expected from Figure~\ref{f:2dmotions}, the orbital pole of NGC~147 
is found to be remarkably close to the normal of the GPoA. This provides 
support that NGC~147 is rotating within the GPoA. The orbital pole of 
NGC~185 is offset by $\sim 16$\degr\ in angle from the GPoA normal, 
but it is still consistent with being dynamically associated with the 
GPoA within $1\sigma$ uncertainty. Improved PM measurements for NGC~185 
(and for M31) are necessary for conclusively judging the association.

In contrast to NGC~147 and NGC~185, the two other satellites shown in
Figure~\ref{f:orbpoles}, M33 and IC~10, are significantly offset from 
the GPoA. M33 is not part of the GPoA originally identified by 
\citet{iba13} and \citet{con13}, so it is unsurprising that its 
orbital pole is so far removed from that of the GPoA. IC~10, however, 
spatially lies within the GPoA though it too was not originally 
identified as a coplanar satellite based on its $\vlos$. IC10's large 
tangential motion with respect to the GPoA indeed indicates that it 
is most likely not dynamically associated with the GPoA.

With our 3-d orbits, we can also assess whether the orbital poles of 
NGC~147 and NGC~185 deviate from their present-day values to determine 
if they have been long-term GPoA members, assuming that the GPoA is not 
a transient structure (see Section~\ref{ss:testgpoa}). To do so, we 
calculate $\theta_\mathrm{disk,sat}$ defined as the angle between the 
angular momentum vectors of NGC~147/185 (computed from the positions 
and velocities of each time step from the orbits reported in 
Section~\ref{s:orbhist}) and the M31 disk plane for both M31 $\vtan$ 
zero points. We find only small variations of $\Delta \theta_\mathrm{disk,sat} 
\sim 1\degr$ as the satellites orbit around M31. For NGC~147 and 
NGC~185, respectively, we find $\theta_\mathrm{disk,sat} = 57\pm9\degr$ 
and $66\pm14\degr$ using the \hst+sats M31 $\vtan$. For the \hst+\gaia\ 
M31 $\vtan$, $\theta_\mathrm{disk,sat} = 63\pm8\degr$ and $74\pm9\degr$
for NGC~147 and NGC~185, respectively. All calculations assume a 
low-mass M31, but results are consistent for a high-mass M31. 

The GPoA is tilted $50.5\degr$ relative to the normal of M31's disk 
with an uncertainty of approximately $10.9\degr$\footnote{This 
uncertainty represents the rms scatter in the angular offsets out 
of the GPoA plane for the 19 satellites that define the GPoA in 
\citet[][see Table 4]{paw13}. The angular offset is defined as 
the arctangent of (perpendicular offset from GPoA / radial distance 
from M31).}. For both satellites, the \hst+\gaia\ M31 $\vtan$ results 
in larger offsets from the GPoA. The resulting $\theta_\mathrm{disk,sat}$ 
for NGC~147 suggests that its motion is generally consistent with 
the orientation of the GPoA to within $\sim10\degr$ and either M31 
$\vtan$ zero point. NGC~185 is consistent with the GPoA and its 
uncertainty for the \hst+sats M31 $\vtan$ zero point, however for 
the \hst+\gaia\ zero point, NGC~185 can be off from the GPoA by up to 
$\sim33\degr$ over the last 6~Gyr when considering uncertainties in 
the observed parameters. Thus, it is possible that NGC~185 is not 
dynamically associated to the GPoA either now or in the past and 
improved PMs for both M31 and the satellites are necessary to 
conclusively determine if NGC~185 is associated to the GPoA. These 
results also reinforce that the two galaxies are not strongly 
interacting, as their orbits do not indicate binary motion about 
a common center of mass.

Since our orbital calculations assume a spherical halo, it is expected 
that the satellites orbit approximately in a plane with a constant 
angle with respect to M31's disk. As the satellites only complete 
1--1.5 orbits around M31 in this time frame, it is unsurprising 
that these angles do not significantly change with time. Additionally, 
these calculations do not include the effects of other dynamical 
perturbers that may have influenced the past orbits of NGC~147 and 
NGC~185, such as M33, M32, or the progenitor of the Giant 
Southern Stream.  

\subsection{Testing the GPoA as a Long-lived Structure}
\label{ss:testgpoa}

The GPoA and the corresponding satellite planes observed around 
the MW have given rise to an extensive literature assessing their consistency
with current cosmological ideas \citep[see][and references therein]{paw19}.  
Different authors find varying levels of consistency with cosmological 
simulations depending on simulation and analysis methods, though in 
general there is a significant level of tension. Some studies find better 
agreement by means of baryonic effects not implemented in most 
simulations \citep[e.g.,][]{saw16}, while others argue that the 
planes are tidal debris structures forming in models outside the 
$\Lambda$CDM paradigm \citep[e.g.,][]{ham13}. In most studies, however, 
the simulated planes are not highly coherent, and therefore claimed as 
chance alignments rather than long-lived structures \citep{bah14,buc15,gil15,san20}. 
Here we simplify the range of models into two extreme possibilities: 
entirely chance alignments, or long-lived coherently rotating structures.
In the former case, one would expect the random motions normal to the 
plane to be of order the overall satellite velocity dispersion of 
$\sigma_v \approx 100$~$\kms$ \citep{tol12,gil18}. In the latter, 
these motions should be $\ll \sigma_v$. \citet{fer17} find the velocity 
dispersion in a long-lived plane structure should be $<20$~$\kms$, and 
given the size of our observational errors we simply take it to be zero. 

The small out-of-plane velocities we have found are, of course, 
{\em consistent} with both models, since these two satellites could 
just happen by chance to be among the few moving within the plane.
We can instead ask how much the new PM results evidence 
{\em strengthens} the case for the GPoA as a long-lived structure. 
From a Bayesian point of view, this requires evaluating the likelihood
function of the new observations in each model (here called 
``transient'' and ``stable'' models as shorthands).

We use the fact that the plane is viewed almost edge-on. Scaling PM to 
distance, the observable quantity is the M31-motion-corrected 
tangential velocity in the direction of the GPoA plane normal vector, 
$v_\mathrm{normal}$. We assume Gaussian distributions in both 
observational errors and physical velocity. For each model the mean of 
$v_\mathrm{normal}$'s distribution is zero. For the stable model the 
likelihood function is then a Gaussian distribution 
$\mathcal{N}(v_\mathrm{normal} \, | \, 0, \,  \sigma_\mathrm{obs}^2 + \sigma_\mathrm{ref}^2)$, 
where the terms in the variance come from the proper motion errors in 
our observations and in the reference frame. In the transient model the 
likelihood function is 
$\mathcal{N}(v_\mathrm{normal} | 0, \sigma_\mathrm{obs}^2 + \sigma_\mathrm{ref}^2 + \sigma_v^2)$,
where the extra term in the variance comes from the random physical velocities.

For NGC~147, the velocity components in W and N directions,
corrected for the M31 system motion using the \hst+sats M31 $\vtan$,
are 38 and 161~$\kms$. This yields an observed velocity of 
$v_\mathrm{normal} =2$~$\kms$ out of the plane. The observed PM 
uncertainty of 49~$\kms$ combined with the reference frame uncertainty 
of 30~$\kms$ yields a total expected dispersion of 57~$\kms$.
Including the expected satellite velocity dispersion $\sigma_v$ 
raises the expected observed dispersion in the transient model to 
115~$\kms$, twice as high. Evaluating the likelihood functions 
above gives a likelihood ratio of 2.0; i.e., the stable model 
yields twice the likelihood of the transient model. 

For the case of NGC~185, velocity components in M31's frame in the 
W and N directions are 49 and 52~$\kms$, which yields an observed 
velocity of $v_\mathrm{normal} =38$~$\kms$ out of the plane. 
The observed PM uncertainty of 43~$\kms$ combined with the reference 
frame uncertainty yields a total expected dispersion of 52~$\kms$. 
The total expected dispersion in the transient model is 113~$\kms$, 
and the resulting likelihood ratio is 1.8. 

Combining the factors for each galaxy gives a total likelihood
ratio of 3.5. Therefore, regardless of one's assessment on the 
stability (or even existence) of the plane of satellites in M31, our 
PM observations strengthen the evidence for the stable model as 
opposed to the transient model by nearly a factor of 4.
This is {\em in addition} to all prior evidence for the existence of 
the plane, based on the sky positions and $\vlos$ of the satellites.

Adopting the \hst+\gaia\ zero-point for M31 $\vtan$ shifts the 
out-of-plane velocities of both galaxies somewhat. For NGC~147 
$v_\mathrm{normal}$ increases to 43~$\kms$, while for NGC~147 it 
increases to 78~$\kms$. As a result, the likelihood ratios change 
from 2.0 to 1.6 for NGC~147 and from 1.8 to 0.9 for NGC~185. 
Combining the two results reduces the overall likelihood ratio from 
3.5 to 1.4. In other words, the new PM observations with this reference 
frame choice are still positive evidence for the stable model, but not 
as strong as with \hst+sats. We anticipate that the uncertainty in 
M31's PM will decrease in the near future, and at that time it will be 
worth revisiting the various calculations we have made that are 
sensitive to M31's reference frame.

\section{Conclusions}
\label{s:conclusions}

We used \hst\ ACS/WFC and WFC3/UVIS images to measure the first PMs of 
NGC~147 and NGC~185, two dwarf satellites of M31. By comparing the 
average motions from a few thousands to tens of thousands stars 
associated with the dwarf galaxies against hundreds of compact 
background galaxies in the fields, we find the PMs to be: 
$(\muw, \mun)_\mathrm{N147} = (-0.0232, 0.0378) \pm (0.0143, 0.0146) \masyr$ 
and $(\muw, \mun)_\mathrm{N185} = (-0.0242, 0.0058) \pm (0.0141, 0.0147) \masyr$.

We derive space velocities of NGC~147 and NGC~185 in the M31-centric frame. 
For the rest-frame M31-centric 
3-d velocities, we use two different $\vtan$ measurements for M31, 
the average of \hst\ measurement and satellite kinematics by \citet{vdm12}, 
and the average of \hst\ and \gaia~DR2 measurements by \citet{vdm19}. 
Using escape velocities based on the measured 3-d velocities, we find 
that M31 must have a minimal virial mass of $0.5-1\times 10^{12}~\Msun$ for 
both NGC~147 and NGC~185 to be bound to M31's halo simultaneously.

We used the PM results to explore the orbital histories of NGC~147 and 
NGC~185 over the past 6 Gyr. These orbital histories account for the 
full 3-body encounter between M31+NGC~147+NGC~185 and explore two M31 
mass combinations, as well as two M31 $\vtan$ zero points. We conclude 
that NGC~147's tidal tails are likely the remnants of a recent 
pericentric passage around M31 at a distance of $\sim64$--$70$~kpc just 
$0.3$--$0.5$~Gyr ago. The direction of motion for NGC~147 is also spatially 
well aligned with the tidal tails. The lack of tidal tails around NGC~185 
is consistent with the orbital results, which indicate that NGC~185 either 
passed around M31 at earlier times (1.5--2 Gyr ago) and that these tidal 
tails are no longer visible today or that NGC~185 passed nearby M31 at very 
early times ($>$ 6 Gyr ago) and at larger distances ($>200$ kpc) such 
that no tidal features ever existed. Future $N$-body models will  
help confirming (or refuting) such claims. The disparities in gas 
content and SFHs found for NGC~147 and NGC~185 can be attributed to 
the fact that the two galaxies have different orbital characteristics.

The orbital histories of NGC~147 and NGC~185 also allow us to revisit 
the claim that these two satellites are a bound pair. Based on the orbital 
histories calculated in this work, we find that the minimum separation 
between the two galaxies was only achieved in the last $\sim$0.5 Gyr 
at a distance of $\sim$100 kpc, thus the satellites show no evidence 
for a strong interaction in the past such that they might have been 
bound to each other. This also implies the satellites may not have been 
born in similar environments, despite their similarities, and furthermore 
that they did not enter the M31's halo as a pair.

Finally, our results provide a new test of the nature of the GPoA. 
For long-term association of a satellite with this plane, its orbital
angular momentum must be perpendicular to the plane (i.e., orbital 
motion within the plane). Our PM measurements yield the first determination 
of the 3-d angular momentum vector of any of the M31 satellites originally 
identified as GPoA members. The orbital pole of NGC~147 agrees within the 
normal direction of the GPoA. The same is true for NGC~185 when using 
the \hst+\gaia\ M31 $\vtan$ zero point, but  with larger uncertainties. 
While additional samples and improved PM uncertainties are required for 
more stringent tests, our results significantly strengthen the hypothesis 
that the GPoA is a dynamically coherent entity.

In a separate future work, we will use the orbital histories of NGC~147 
and NGC~185 derived from this work to provide insight into the formation 
and evolution histories of dE satellites using dark matter-only 
cosmological simulations. 

Whereas the increased time baselines for subsequent \gaia\ data releases 
will undoubtedly yield enhanced understanding about the formation and 
evolution of the MW system, PM measurements at the distance of M31 will 
require deep imaging using space telescopes like \hst, \textit{JWST}, and 
\textit{WFIRST}. Fortunately, there are a number of existing and ongoing 
(GO-15902; PI: D.~R. Weisz) \hst\ programs targeting virtually all known 
M31 satellites to measure their SFHs. These can be used as first-epoch data 
for PM measurements with follow-up by additional \hst\ (or even 
\textit{JWST}/\textit{WFIRST}) observations. Additionally, the refined 
M31 PM measurements from the ongoing \hst\ program GO-15658 (PI: 
S.~T. Sohn) will provide improved zero points for the satellites' 
tangential velocities. In this regard, prospects of M31 satellite 
dynamics studies are very bright.

\acknowledgments
Support for this work was provided by NASA through grants for program 
GO-14769 from the Space Telescope Science Institute (STScI), which is 
operated by the Association of Universities for Research in Astronomy 
(AURA), Inc., under NASA contract NAS5-26555. E.P. is supported by 
the Miller Institute for Basic Research in Science at UC Berkeley.
The authors wishes to thank M. Pawlowski for providing helpful comments 
about the plane of satellites. We also thank the referee for providing 
useful comments that improved the quality of this paper.

\facility{HST (ACS/WFC and WFC3/UVIS)}

\appendix

\section{Correlations between Orbital Parameters}
\label{s:app1}

\begin{figure*}
\centering
\vspace{-2mm}
\includegraphics[width=0.9\textwidth]{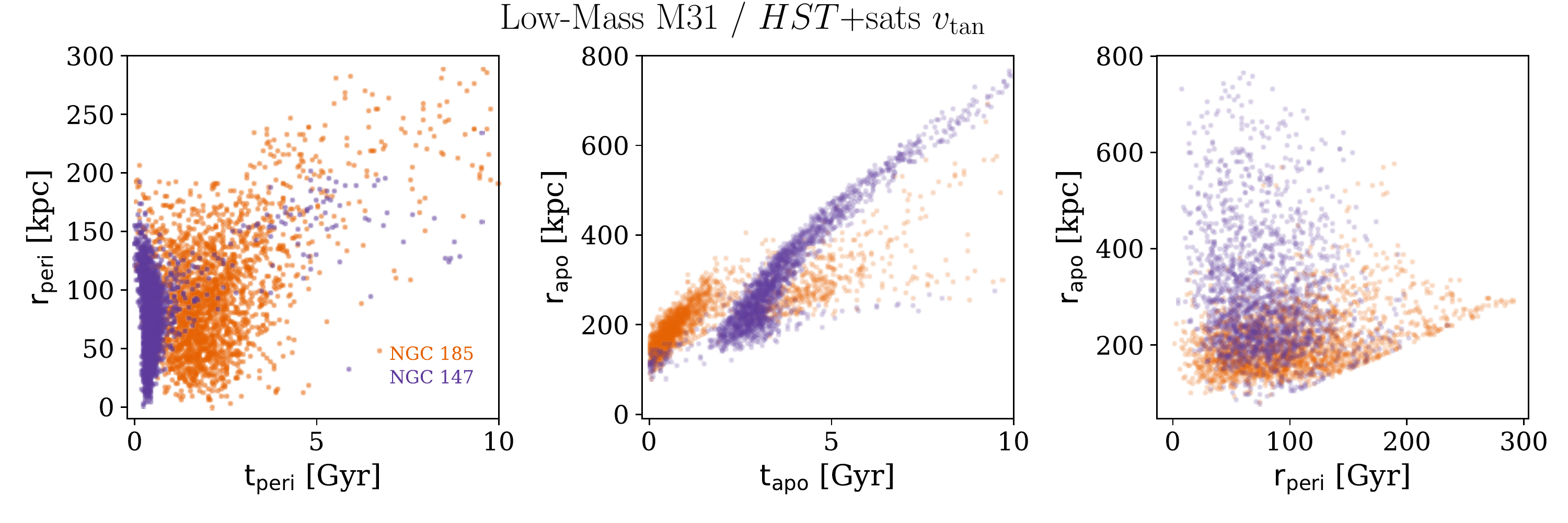}\\
\vspace{-2mm}
\includegraphics[width=0.9\textwidth]{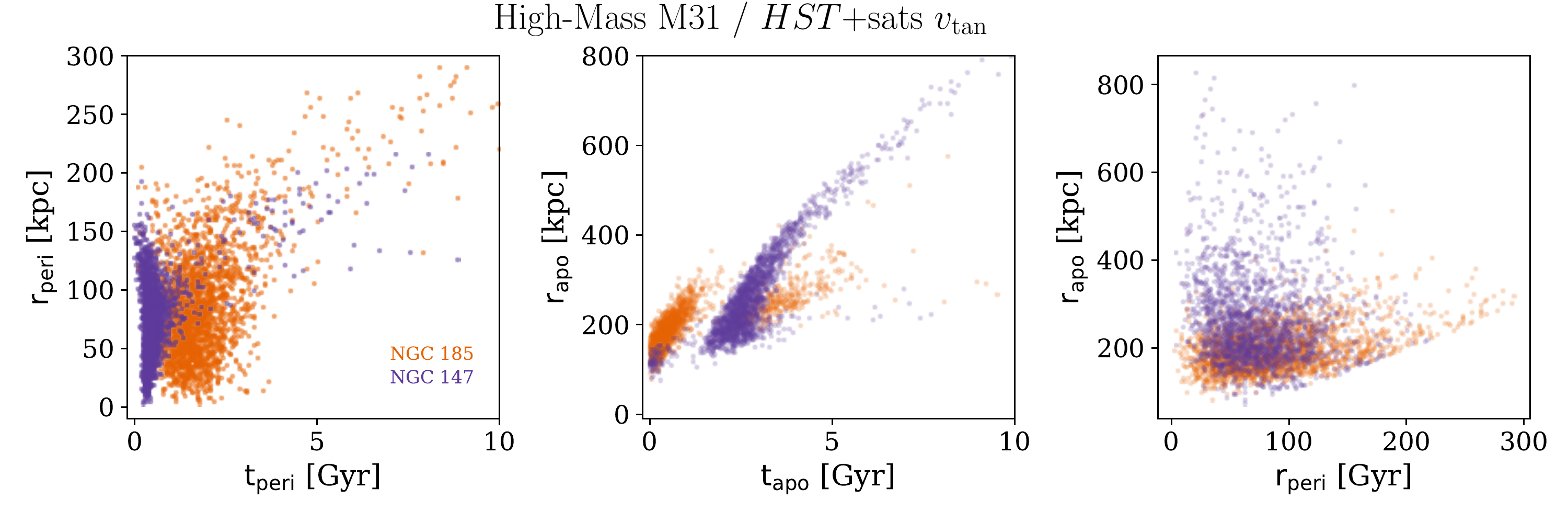}\\
\vspace{-2mm}
\includegraphics[width=0.9\textwidth]{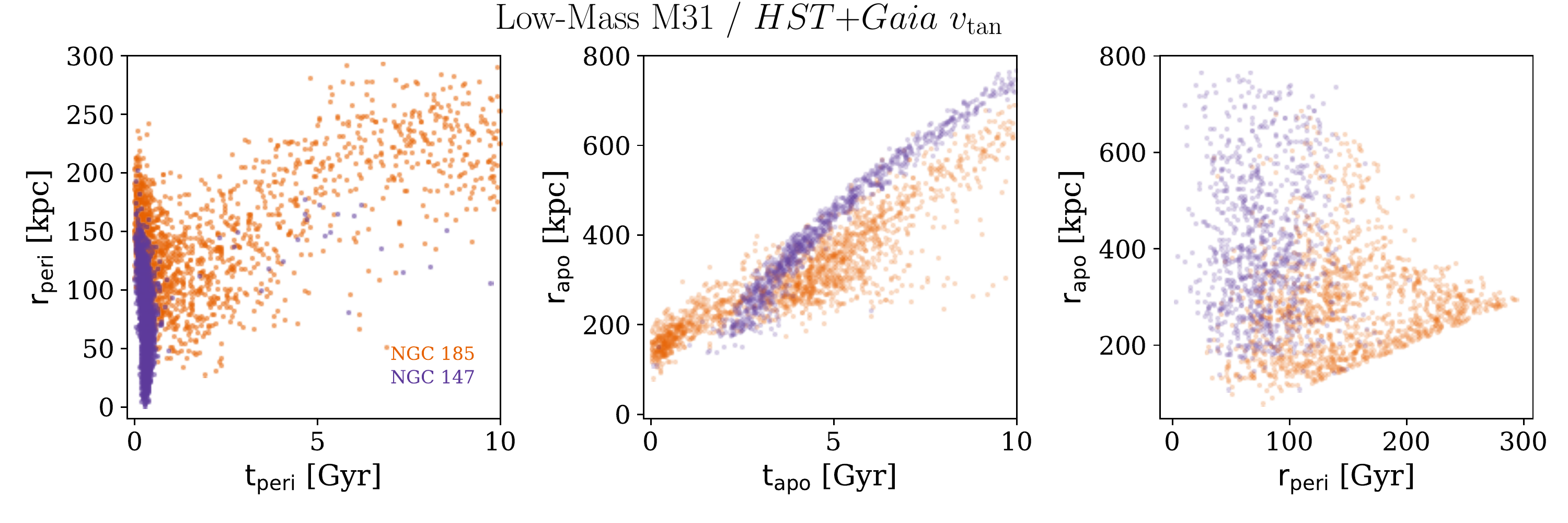}\\
\vspace{-2mm}
\includegraphics[width=0.9\textwidth]{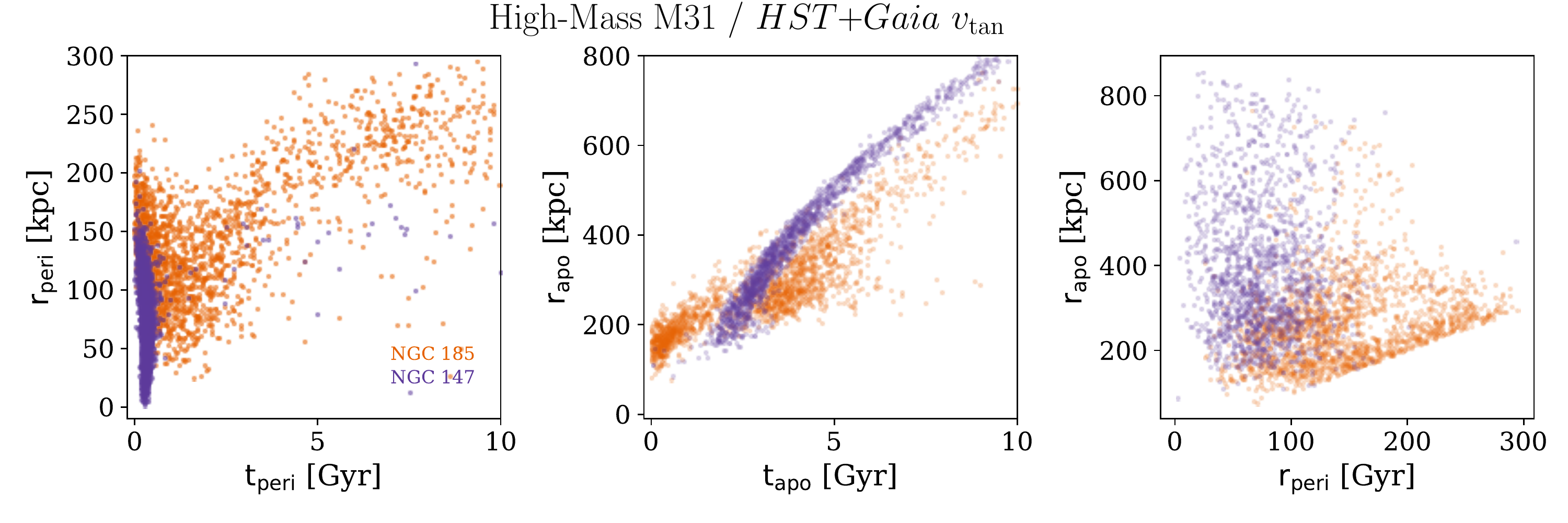}
\caption{Correlations of peri- and apo-centric parameters resulting from propagating 
         the observed parameters in a Monte Carlo fashion. From left to right, 
         $\rperi$ vs. $\tperi$, $\rapo$ vs. $\tapo$, and $\rperi$ vs. $\rapo$ 
         are shown for different combinations of M31 mass and $\vtan$ adopted 
         in our study. Each data point in these plots represent a random sampling 
         drawn from our propagation of observed uncertainties. Note that a 
         pericentric passage and/or an apocentric passage are not recovered in 
         every individual orbit resulting from the random sampling. Columns   
         $f_\mathrm{peri}$ and $f_\mathrm{apo}$ in Table~\ref{t:orbparams} list 
         the percent of 2,000 orbits where these orbital extrema are found. 
         These percentages correspond to the number of data points plotted in 
         each panel. The right column contains the fewest points in each row since 
         the percentage of orbits that complete at least half of an orbital period 
         (i.e. both a pericenter and apocenter) is always fewer than the percentage 
         of orbits where either an apocenter or pericenter are recovered.
         \label{f:orbparams_corr}}
\end{figure*}

In Figure~\ref{f:orbparams_corr}, we show the correlations of peri- and 
apo-centric parameters from the samplings drawn from our propagation 
of observed uncertainties.

%
%

\section{Uncertainties in Orbital Parameters}
\label{s:app2}

\begin{figure*}
\includegraphics[width=1\textwidth]{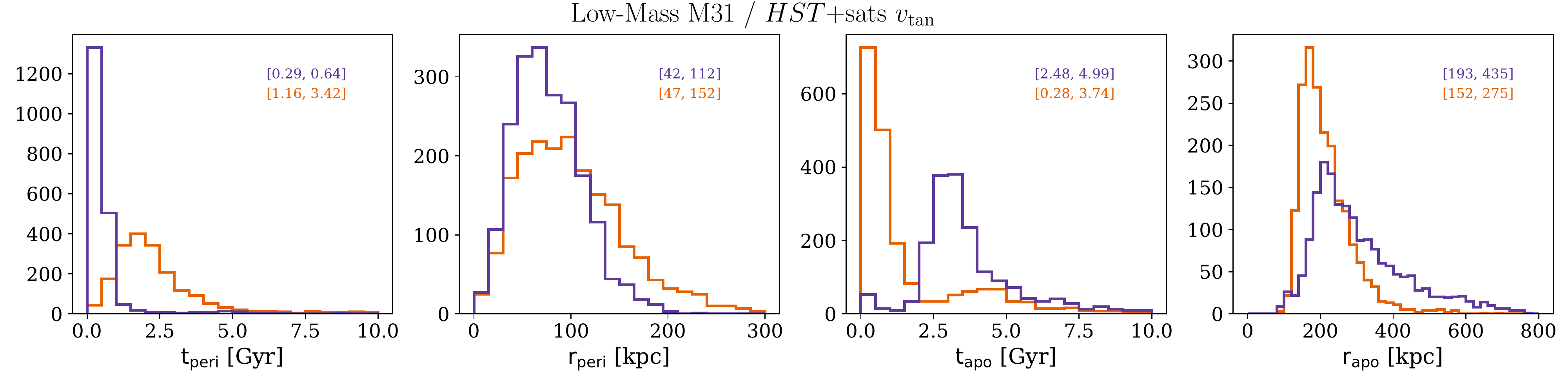}\\
\includegraphics[width=1\textwidth]{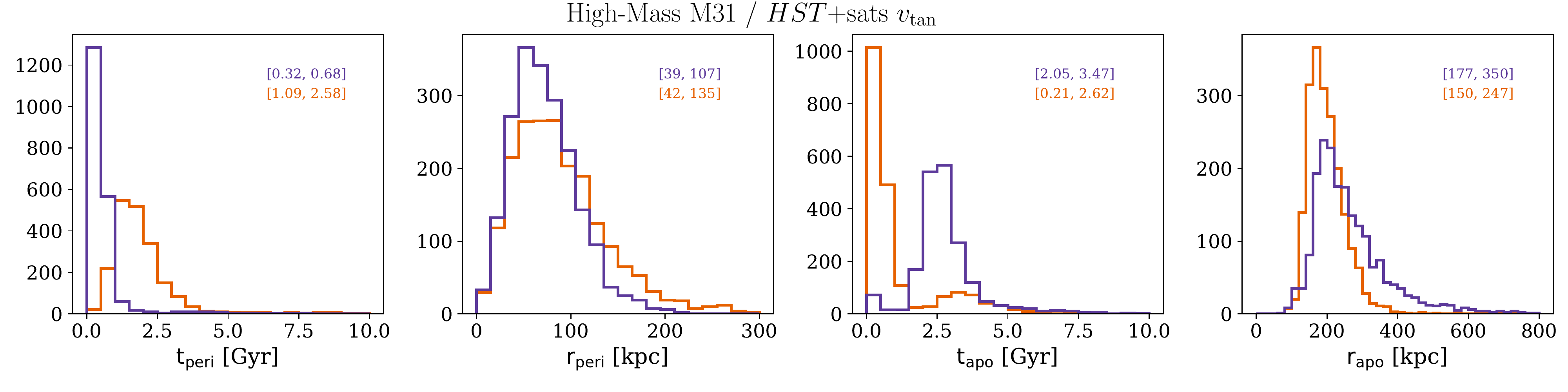}\\
\includegraphics[width=1\textwidth]{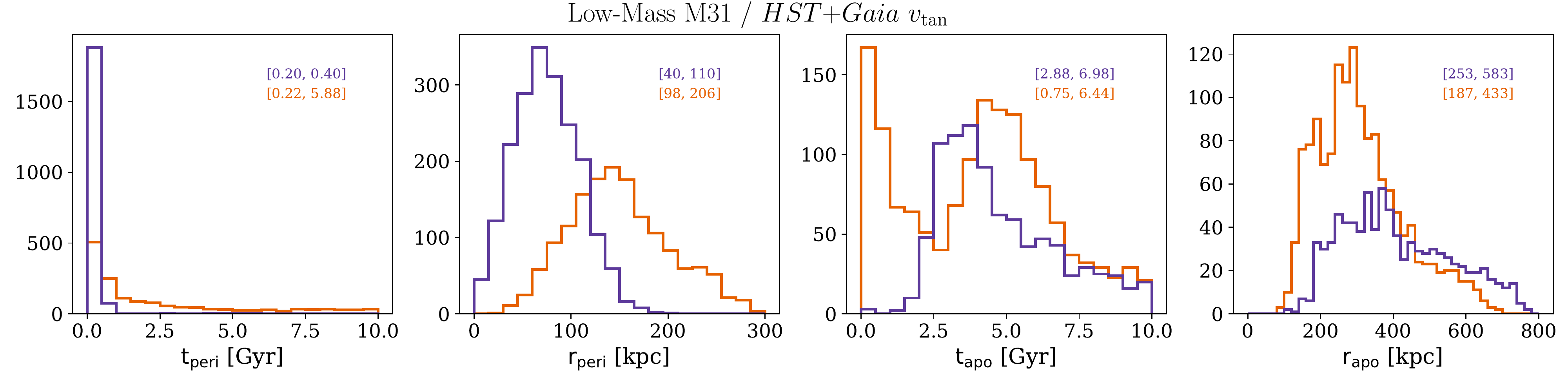}\\
\includegraphics[width=1\textwidth]{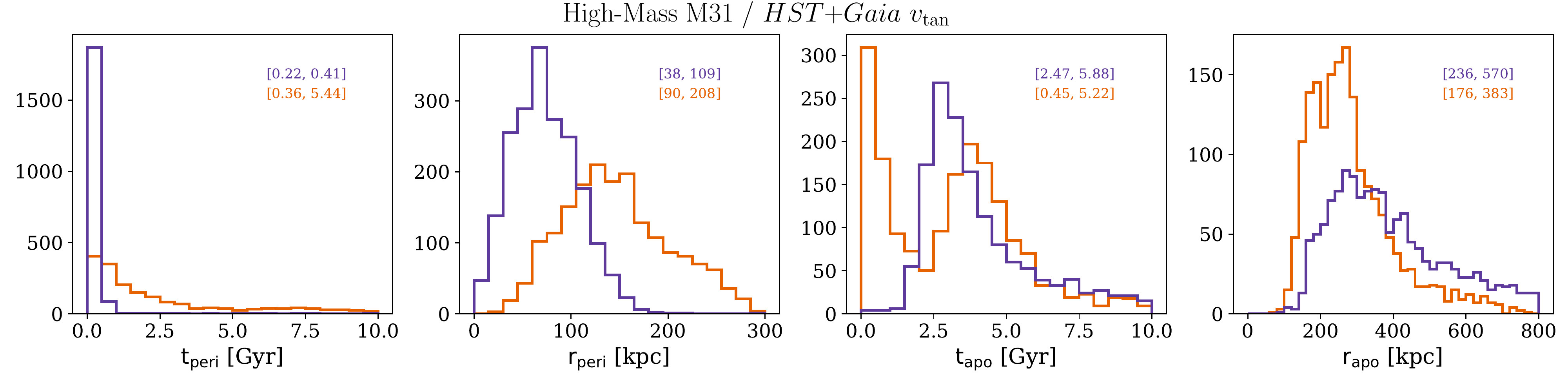}
\caption{The distributions of peri-/apo-centric times and distances 
         resulting from propagating the observed parameters in a Monte Carlo 
         fashion using four different combinations of M31 mass and $\vtan$ adopted 
         in our study. The same color schemes as in our previous plots are used
         (purple for NGC~147, and orange for NGC~185). On the top right of 
         each panel, we list the $[15.9, 84.1]$ percentiles around the median of the distributions.
         \label{f:orbparams_hist}}
\end{figure*}

In Figure~\ref{f:orbparams_hist}, we show the distributions of 
peri-/apo-centric times and distances resulting from propagating the 
observed parameters in a Monte Carlo fashion.

%
%



\end{document}